\documentclass[12pt]{article}
\usepackage{epsfig}

\usepackage{graphicx}
\usepackage{cancel}
\usepackage{amssymb,amsmath}

\DeclareMathOperator\arctanh{arctanh}

\def\harr#1#2{\smash{\mathop{\hbox to .3in{\rightarrowfill}}
 \limits^{\scriptstyle#1}_{\scriptstyle#2}}}

\def\s2{\frac{1}{\sqrt2}}

\def\beqa{\begin{eqnarray}}
\def\eeqa{\end{eqnarray}}

\def\Dsl{\,\raise.15ex\hbox{/}\mkern-13.5mu D} 
\def\d3{d^3}

\newcommand{\be}{\begin{equation}}
\newcommand{\ee}{\end{equation}}
\newcommand{\beq}{\begin{eqnarray}}
\newcommand{\eeq}{\end{eqnarray}}

\renewcommand{\Im}{\text{Im}\,}

\usepackage{amssymb,amsmath}

\usepackage{hyperref}
\hypersetup{
        colorlinks=true,
        linkcolor=blue,
        citecolor=red,
}

\def\be{\begin{equation}}
\def\ee{\end{equation}}
\def\beqa{\begin{eqnarray}}
\def\eeqa{\end{eqnarray}}

\begin{document}

\begin{center}
\Large{\bf Lorentzian Vacuum Transitions for Anisotropic Universes}
\vspace{0.5cm}

\large  H. Garc\'{\i}a-Compe\'an\footnote{e-mail address: {\tt
compean@fis.cinvestav.mx}}, D. Mata-Pacheco\footnote{e-mail
address: {\tt dmata@fis.cinvestav.mx}}

\vspace{0.3cm}

{\small \em Departamento de F\'{\i}sica, Centro de
Investigaci\'on y de Estudios Avanzados del IPN}\\
{\small\em P.O. Box 14-740, CP. 07000, Ciudad de M\'exico, M\'exico}\\

\vspace*{1.5cm}
\end{center}

\begin{abstract}
The vacuum transition probabilities for anisotropic universes in the
presence of a scalar field potential in the Wentzel-Kramers-Brillouin approximation are
studied. We follow the work by Cespedes \textit{et al}
[Phys. Rev. D 104, 026013 (2021)], which discuss these transitions in the
isotropic context using the Wheeler-DeWitt equation, the
Lorentzian Hamiltonian approach and the {\it thin wall} limit.
First, we propose a general procedure to adapt their formalism to
compute the decay rates for any superspace model. Then we apply it
to compute the transition probabilities of an Friedmann-Lemaitre-Robertson-Walker (FLRW) metric with both
positive and zero curvature, reproducing in this way one of the
results obtained at Cespedes \textit{et al}. We then proceed to
apply the formalism to three anisotropic metrics, namely,
Kantowski-Sachs, Bianchi III and biaxial Bianchi IX to compute the
rate decays for these three cases. In the process we find that this
method involves some conditions which relates the effective number
of independent degrees of freedom resulting on all probabilities
being described with only two independent variables. For the Bianchi III metric, we find that a general effect of anisotropy is to decrease the transition probability as the degree of anisotropy is increased, having as the isotropic limit the flat FLRW result.

\vskip 1truecm

\end{abstract}

\bigskip

\newpage

\section{Introduction}
\label{sec:intro}

For some decades the theory of quantum gravity has been one of the
most important challenges of theoretical physics. Over the years,
many approaches to this subject have been developed presenting a
list of different efforts to achieve a consistent theory. One of the
first attempts in this regard is the use of a Hamiltonian formalism
and canonical quantization \cite{Arnowitt:1962hi} to obtain a
quantum picture of gravity leading to a Schr\"odinger-like equation,
namely, the Wheeler-DeWitt (WDW) equation \cite{Wheeler,DeWitt}. In
this formalism there is not a time variable in which the wave
functionals can depend on, leading to difficulties when interpreting
the solutions (the problem of time). Nevertheless, the decay rate
i.e. the ratio of absolute squares of the wave functionals for
different configurations can be interpreted as relative
probabilities between two states in which the system performs a
transition.

Cosmological problems of the early universe are a natural scenario
in which the effects of such a quantum theory are expected to be of
first importance. In particular, there has been great interest over
the years in the study of effects of vacuum decay in this type of
theories. As in field theory, this process is described by the
nucleation of true vacuum bubbles. Euclidean methods to study these
processes were developed originally by Sidney Coleman and
collaborators for field theory \cite{ColemanFT1,ColemanFT2}  and
then, they were extended to include gravity by Coleman and De Luccia
\cite{ColemanDeLuccia}. Their results were later generalized by
Parke \cite{Parke:1982pm}. Recently a Hamiltonian approach to study
vacuum transitions without using Euclidean methods was developed in
Refs. \cite{FMP1,FMP2}. The Fischler, Morgan and Polchinski (FMP)
procedure was used in \cite{deAlwis:2019dkc,Cespedes:2020xpn} to
compute tunneling probabilities of transitions between two de
Sitter universes and the transition from Minkowski to de Sitter is
also possible without having singular instantons. A generalization
of FMP method to include explicitly scalar field potentials for the
general Wheeler's superpace  and using a Wentzel-Kramers-Brillouin
(WKB) approximation was developed in the work
\cite{Cespedes:2020xpn}. Although the method can be defined on the
general superspace, in order to compute some concrete observable,
for simplicity one usually is restricted to minisuperspace. The
method is applicable to any number of fields and with the additional
{\it thin wall} limit, it produces the same results for a closed
Friedmann-Lemaitre-Robertson-Walker (FLRW) metric as the one
obtained by the uses of Euclidean methods
\cite{deAlwis:2019dkc,Cespedes:2020xpn}.

On the other hand, many different studies based in satellite
observations have been carried out regarding the deviations of the
cosmological isotropy. For instance analysis from the Cosmic
Microwave Background (CMB) measures of temperature and polarization
from Planck observatory \cite{Saadeh:2016sak}, seems to strongly
imply that the deviations from isotropy are disfavored. However
other observations on type Ia supernovae of the cosmic acceleration
\cite{Colin:2018ghy} seems to imply certain degree of anisotropy
which may play a fundamental role in the cosmological models.
Although a more recent work on this subject challenges this
assertion and gives support to statistical isotropy
\cite{Soltis:2019ryf}. Other groups have studied the anisotropy
using based-space x-ray observations and have found some results
compatible with a nonvanishing degree of large scale anisotropy
\cite{Migkas:2020fza}. Moreover some implications of the anisotropy
in galaxy clusters is analyzed in \cite{Migkas:2021zdo}. More
recently, the study of the tension in the value of the Hubble
constant would imply a modification of the current paradigm Einstein
theory + FLRW Universe. The presence of anisotropies seem to be
compatible with those results
\cite{Krishnan:2021dyb,Krishnan:2021jmh}. Then the debate about the
possible existence of a certain degree of anisotropy is still not
conclusive when all observations are taken into account. Thus the
importance of studying homogeneous but anisotropic universes in the
context of quantum gravity is quite relevant and not empty. Vacuum
transitions for these types of universes have also been studied
previously. In reference \cite{DelCampo}, it was described these
transitions involving only gravity and a cosmological constant for
the Bianchi IX model. In the mentioned work, these transitions have
also been studied in the presence of a scalar field. Moreover, in
Ref. \cite{JensenRuback}, it is described these transitions for a
Kantowski-Sachs metric using also Euclidean methods and in reference
\cite{Folomeev}, the Bianchi IX model is studied in the tunneling
from nothing scenario. Finally, in reference \cite{Mansouri}
transitions for two minima of a scalar field potential for the
Bianchi I model was also studied into this context.

In the present article we study vacuum transitions using the
Hamiltonian formalism developed in \cite{Cespedes:2020xpn} for three
anisotropic metrics: $(a)$ Kantowski-Sachs, $(b)$ Bianchi III and
$(c)$ the biaxial Bianchi IX. In particular, we focus on obtaining
transition probabilities in the semiclassical approximation for two
minima of the potential in a Lorentzian computation, without
resorting to any Euclidean method.

This work is organized as follows. In Section \ref{S-WKB} we give a
brief review of the general proposal of reference
\cite{Cespedes:2020xpn}. It is also presented a general procedure to
apply this method to any superspace model. In the next sections the
procedure is applied to some examples in the minisuperspace. Section
\ref{S-FLRW} is devoted to apply our method to the FLRW metric
with positive curvature and we show that our procedure gives the
same results as the original work, we also point out a particular
remark. We apply our method to obtain the transition probabilities
in the remaining sections for the FLRW isotropic and plane metric in
Section \ref{S-FLRWP} and for the anisotropic metrics:
Kantowski-Sachs in Section \ref{S-KS}, Bianchi III in Section
\ref{S-B3} and biaxial Bianchi IX in Section \ref{S-B9}. Finally, in
Section \ref{S-Dis} we discuss our results.

\section{WKB approximation for quantum gravity}
\label{S-WKB}

In this section we start by introducing the general approach
proposed in Ref. \cite{Cespedes:2020xpn}, which applies the WKB
approximation to the WDW equation. We follow the notation and
conventions given at Ref. \cite{Cespedes:2020xpn}. In particular, we
stress our attention in obtaining a general expression for the
semiclassical decay rate between a false and a true vacuum of a
scalar field potential.

In the context of the ADM-formulation of general relativity (GR)
\cite{Arnowitt:1962hi,Wheeler,DeWitt} we know that the general form
of the Hamiltonian constraint is given by
    \begin{equation}\label{HamConstraint}
        \mathcal{H}=\frac{1}{2}G^{MN}(\Phi)\pi_{M}\pi_{N}+f[\Phi]\approx0 ,
    \end{equation}
where $G^{MN}$ is the inverse metric in Wheeler's superspace, $\Phi$
represents all the coordinates $\Phi^M$ of the superspace (which can
be infinite dimensional), that is, the components of the
three-dimensional metric, the matter field variables, etc. and their
corresponding canonically conjugate momenta $\pi_{M}$. Moreover
$f[\Phi]$ stands for all the other terms of 3-curvature and
potential terms of matter fields that arise in the WDW equation. The
WDW equation is obtained after doing a canonical quantization of the
Hamiltonian constraint, that is, replacing
$\pi_{M}\to-i\hbar\frac{\delta}{\delta\Phi^M}$ in the last
expression, from where we obtain up to ordering ambiguities
    \begin{equation}\label{WDWEQ}
        \mathcal{H}\Psi(\Phi)=\left[-\frac{\hbar^2}{2}G^{MN}(\Phi)\frac{\delta}{\delta\Phi^M}\frac{\delta}{\delta\Phi^N}+f[\Phi]\right]\Psi[\Phi]=0 ,
    \end{equation}
where $\Psi[\Phi]$ is the usual wave functional in superspace.

In order to obtain a semiclassical result, we take the general
proposal of the WKB type
    \begin{equation}\label{WKBProposal}
        \Psi[\Phi]=\exp\left\{\frac{i}{\hbar}S[\Phi]\right\} ,
    \end{equation}
with the $\hbar$-expansion
    \begin{equation}\label{SemiClasExp}
        S[\Phi]=S_{0}[\Phi]+\hbar S_{1}[\Phi]+\mathcal{O}(\hbar^2) .
    \end{equation}
Substituting this in the WDW equation (\ref{WDWEQ}) we obtain for
the two lowest order in $\hbar$
    \begin{equation}\label{WDWOrder0}
            \frac{1}{2}G^{MN}\frac{\delta S_{0}}{\delta\Phi^{M}}\frac{\delta S_{0}}{\delta\Phi^{N}}+f[\Phi]= 0 ,
    \end{equation}
    \begin{equation}\label{WDWOrder1}
        2G^{MN}\frac{\delta S_{0}}{\delta \Phi^M}\frac{\delta S_{1}}{\delta\Phi^N}=iG^{MN}\frac{\delta^2}{\delta\Phi^M\delta\Phi^N}S_{0} .
    \end{equation}
Given a set of integral curves with parameter $s$ defined on a
particular slice on the space of fields, they are described by
    \begin{equation}\label{DefCs}
        C(s)\frac{d\Phi^M}{ds}=G^{MN}\frac{\delta S_{0}}{\delta\Phi^N} .
    \end{equation}
We can write down the classical action in the form
    \begin{equation}\label{ClassAction0}
        S_{0}[\Phi_{s}]=\int^{\Phi_{s}}\int_{X}\pi_{M}dx^{M} ,
    \end{equation}
which,  by using Eqs. (\ref{WDWOrder0}) and (\ref{DefCs}), it can be
rewritten as
    \begin{equation}\label{ClassActionGen}
        S_{0}[\Phi_{s}]=-2\int^s\frac{ds'}{C(s')}\int_{X}f[\Phi_{s'}] .
    \end{equation}
Manipulating Eqs. (\ref{WDWOrder0}) and (\ref{DefCs}) we find the
following relation
    \begin{equation}\label{RelDerC2}
        G_{MN}\frac{d\Phi^M}{ds}\frac{d\Phi^N}{ds}=-\frac{2f[\Phi_{s}]}{C^2(s)} ,
    \end{equation}
where $G_{MN}$ is the inverse of $G^{MN}$.

We can see that Eqs. (\ref{DefCs}) and (\ref{RelDerC2}) form a
system of $n+1$ equations for the $n+1$ variables:
$\left(\frac{d\Phi^M}{ds},C^2(s)\right)$ which in principle can be
solved and then substituted back into Eq. (\ref{ClassActionGen}) to
obtain the classical action. Therefore, in principle, the general
setup presented provides us with enough information to compute the
classical action, and thus the wave functional to first order in
$\hbar$.

Let us assume that the fields $\Phi^M$ depend only on the time
variable, then using Eq. (\ref{ClassActionGen}), the variational
derivative in (\ref{DefCs}) can be expressed in terms of a partial
derivative of the function $f[\Phi]$. By doing this the system can
be solved in general giving as solutions
    \begin{equation}\label{CsGeneral}
        C^2(s)=-\frac{2{\rm Vol}^2(X)}{f[\Phi]}G^{MN}\frac{\partial f}{\partial\Phi^M}\frac{\partial f}{\partial\Phi^N} ,
    \end{equation}
    \begin{equation}\label{DerivSolGen}
            \frac{d\Phi^M}{ds}=\frac{f[\Phi]}{{\rm Vol}(X)}\frac{G^{MN}\frac{\partial f}{\partial\Phi^N}}{G^{LO}\frac{\partial f}{\partial\Phi^L}\frac{\partial f}{\partial\Phi^{O}}} ,
    \end{equation}
where Vol$(X)$ is the volume of the spatial slice. From equation
(\ref{DerivSolGen}) we see that in general, we can obtain the
following system of differential equations relating the various
fields
    \begin{equation}\label{FieldsRelations}
        \frac{d\Phi^M}{d\Phi^N}=\frac{G^{ML}\frac{\partial f}{\partial\Phi^L}}{G^{NP}\frac{\partial f}{\partial\Phi^{P}}} ,
    \end{equation}
which is valid for every value of $M$ and $N$ such that
$d\Phi^{M,N}\neq0$. Therefore, the degrees of freedom are reduced in
general and the classical action will involve less fields on its
computation.

We now consider the matter content of the system to be only a scalar
field canonically coupled to gravity and with a potential that has a
false and a true vacuum. We study a wave functional corresponding to
a path in field space in which the scalar field evolves from the
false minimum to the true one, and one in which the system is kept
at the false minimum during the path.  Then the squared ratio of the
absolute value of these wave functionals can be interpreted as the
transition probability for the system to tunnel from the false to
the true vacuum. Therefore, in the WKB approach considered here, the
transition tunneling probability of going from the false vacuum at
$\phi_{A}$ to the true vacuum at $\phi_{B}$ is the decay rate which
is given by
$$
P(A\to
B)=\left|\frac{\Psi(\varphi^I_{0},\phi_{B};\varphi^I_{m},\phi_{A})}{\Psi(\varphi^I_{0},\phi_{A};\varphi^I_{m},\phi_{A})}\right|^2
$$
\begin{equation}
=\left|\frac{\beta
e^{\frac{i}{\hbar}S_{0}(\varphi^I_{0},\phi_{B};\varphi^I_{m},\phi_{A})}+\chi
e^{-\frac{i}{\hbar}S_{0}(\varphi^I_{0},\phi_{B};\varphi^I_{m},\phi_{A})}}{\beta
e^{\frac{i}{\hbar}S_{0}(\varphi^I_{0},\phi_{A};\varphi^I_{m},\phi_{A})}+\chi
e^{-\frac{i}{\hbar}S_{0}(\varphi^I_{0},\phi_{A};\varphi^I_{m},\phi_{A})}}\right|^2=\left|e^{-\Gamma}\right|^2,
\label{TransProbOr}
\end{equation}
where we have denoted as $\varphi^I$ the remaining fields on
superspace apart from the scalar degree of freedom,
$\Psi(\varphi^I_{0},\phi_{B},\varphi^I_{m},\phi_{A})$ is the wave
functional corresponding to the path that starts in
$\varphi^I(s=0)=\varphi^I_{0}$ with the scalar field  $\phi_{B}$ and
ends in $\varphi^I(s=s_{M})=\varphi^I_{m}$ with the scalar field
$\phi_{A}$. In Eq. (\ref{TransProbOr}) $\beta$ and $\chi$ are the
coefficients of the linear superposition to construct the general
solution for the wave functionals. However, since all the terms
involved are exponentials, we note that in each case one of the
terms would be exponentially bigger than the other, and
consequently, we can safely consider just the dominant term.
Therefore, since we are working on a semiclassical approach, we
will keep only the first term in the WKB expansion
(\ref{SemiClasExp}), $\Gamma$ will be given in any case as
    \begin{equation}\label{DefGamma}
        \pm\Gamma=\frac{i}{\hbar}S_{0}(\varphi^M_{0},\phi_{B};\varphi^M_{m},\phi_{A})-\frac{i}{\hbar}S_{0}(\varphi^M_{0},\phi_{A};\varphi^M_{m},\phi_{A}) ,
    \end{equation}
where the $\pm$ choice depends on which term is dominant in both the numerator and denominator above.
Therefore, we have
    $$
    P(A\to B)=\exp\left[-2{\rm Re}(\Gamma)\right]
    $$
    \begin{equation}\label{TransProb}
    = \exp\left\{\pm2{\rm Re}\left[\frac{i}{\hbar}S_{0}(\varphi^M_{0},\phi_{B};\varphi^M_{m},\phi_{A})-\frac{i}{\hbar}S_{0}(\varphi^M_{0},\phi_{A};\varphi^M_{m},\phi_{A})\right]\right\} .
    \end{equation}

\section{Transitions in FLRW with positive curvature}
\label{S-FLRW}

In the previous section the described formalism was valid in the
Wheeler's superspace which has in general, an infinite number of
variables. However, as it is well known, it is very complicated to
perform a computation in this context. Thus, from now on we will
work in minisuperspace which has a finite number of coordinates.

We start by studying the transitions in the case of a FLRW metric
with positive curvature. The transition probability for this case
was already obtained in \cite{Cespedes:2020xpn} but we derive it
again here using the general method described in the previous
section and we mention a subtlety not considered in that work. The
FLRW metric with positive curvature can be written as
    \begin{equation}\label{MetricFLRWP}
        ds^2=-N^2(t)dt^2+a^2(t)\left(dr^2+\sin^2\theta d\Omega^2_{2}\right) ,
    \end{equation}
where $N(t)$ is the lapse function, $d\Omega^2_{2}$ is the metric of
a $2$-sphere and $0\leq r\leq \pi$. Considering gravity coupled to a
scalar field that depends only on the time variable $\phi=\phi(t)$,
and using natural units in which $c=1$ and $8\pi G$=1 , we obtain
the Lagrangian for this system to be
    \begin{equation}\label{FLRWLagr}
        \mathcal{L}=3N(t)a(t)-\frac{3a(t)\dot{a}(t)^2}{N(t)}+\frac{a^3(t)\dot{\phi}^2}{2N(t)}-N(t)a^3(t)V(\phi) ,
    \end{equation}
where $\dot{a}(t)$ stands for the derivative with respect to the
time variable $t$ and $V(\phi)$ is the scalar field potential.
Performing standard computations we obtain the canonical momenta
    \begin{equation}\label{FLREMomenta}
        \pi_{N}=0 , \hspace{1cm} \pi_{a}=-\frac{6a\dot{a}}{N} , \hspace{1cm} \pi_{\phi}=\frac{a^3\dot{\phi}}{N} ,
    \end{equation}
and  the Hamiltonian constraint
    \begin{equation}\label{FLREHam}
        H=N\left[\frac{\pi^2_{\phi}}{2a^3}-\frac{\pi^2_{a}}{12a}-3a+a^3V\right]\approx
        0.
    \end{equation}
Since the canonical momentum with respect to $N$ vanishes, we can
ignore the prefactor and focus only on the term inside brackets in
the last expression, then by comparing it with the general form of
Eq. (\ref{HamConstraint}) we see that for this metric the set of
variables is $\{\Phi^M\}=\{a,\phi\}$. While the nonvanishing elements
of the metric in minisuperspace are
    \begin{equation}\label{FLREMetricSuper}
        G^{aa}=-\frac{1}{6a} , \hspace{1cm} G^{\phi\phi}=\frac{1}{a^3} ,
    \end{equation}
and
    \begin{equation}\label{FLRWf}
        f(a,\phi)=-3a+a^3V(\phi) .
    \end{equation}
We also note that in this case the volume of the spatial slice is
given by
    \begin{equation}\label{FLRWVolume}
        {\rm Vol}(X)=\int_{\phi=0}^{2\pi}\int_{\theta=0}^{\pi}\int_{r=0}^{\pi}\sin^2r\sin\theta drd\theta
        d\phi=2\pi^2,
    \end{equation}
which is evidently finite.

We choose the parameter $s$ such that for the interval
$[0,\bar{s}-\delta s]$, where $s=0$ is the initial value, the field
remains close to its value at the true minimum $\phi_{B}$, and for
the interval $[\bar{s}+\delta s,s_{m}]$ the field remains very close
to its value at the false minimum $\phi_{A}$, that is, we choose the
parameter $s$ such that
    \begin{equation}\label{ChooseS}
        \phi(s) \approx
        \begin{cases}
        \phi_{B} , & 0<s<\bar{s}-\delta s,\\
        \phi_{A} , & \bar{s}+\delta s<s<s_{M} .
        \end{cases}
    \end{equation}
Therefore, using Eq. (\ref{ClassActionGen}) we have
    \begin{multline}\label{FLRWAccionComplA}
        S_{0}(a_{0},\phi_{B};a_{m},\phi_{A})=-4\pi^2\int_{0}^{s_{M}}\frac{ds}{C(s)}\left[-3a+a^3V(\phi)\right] \\
        =-4\pi^2\bigg\{\int_{0}^{\bar{s}-\delta s}\frac{ds}{C(s)}\left[-3a+a^3V_{B}\right] +\int_{\bar{s}-\delta s}^{\bar{s}+\delta s}\frac{ds}{C(s)}\left[-3a+a^3V(\phi)\right] \\ +\int_{\bar{s}+\delta s}^{s_{M}}\frac{ds}{C(s)}\left[-3a+a^3V_{A}\right]\bigg\} ,
    \end{multline}
where $V_{B}=V(\phi=\phi_{B})$ and $V_{A}=V(\phi=\phi_{A})$. We also
have
    \begin{equation}\label{FLRWAccionComplB}
    S_{0}(a_{0},\phi_{A};a_{m},\phi_{A})=-4\pi^2\int_{0}^{s_{M}}\frac{ds}{C(s)}\left[-3a+a^3V_{A}\right] .
    \end{equation}
Substituting these expressions back in Eq. (\ref{DefGamma}) we
obtain
    \begin{multline}\label{FLRWGammaA}
        \pm\Gamma=-\frac{4\pi^2i}{\hbar}\int_{0}^{\bar{s}-\delta s}\frac{ds}{C(s)}\left[-3a+a^3V_{B}\right]+\frac{4\pi^2i}{\hbar}\int_{0}^{\bar{s}-\delta s}\frac{ds}{C(s)}\left[-3a+a^3V_{A}\right] \\ -\frac{4\pi^2i}{\hbar}\int_{\bar{s}-\delta s}^{\bar{s}+\delta s}\frac{ds}{C(s)}a^3\left[V(\phi)-V_{A}\right].
    \end{multline}
In order to compute the two first integrals, we study the system of
equations (\ref{DefCs}) and (\ref{RelDerC2}) for this metric in the
particular case in which the scalar field is constant. The general
solutions (\ref{CsGeneral}) and (\ref{DerivSolGen}) take in this
case the form
    \begin{equation}\label{FLRWSolutions}
    \begin{split}
            C^{2}(s)&=4\pi^{4}\frac{\left(aV_{A,B}-\frac{1}{a}\right)^{2}}{\frac{a^{2}}{3}V_{A,B}-1} ,
            \\
            \frac{da}{ds}&=\frac{1}{2\pi^{2}}\frac{\frac{a^{2}}{3}V_{A,B}-1}{aV_{A,B}-\frac{1}{a}} ,
    \end{split}
    \end{equation}
where in these solutions $V_{A,B}$ represents either $V_{A}$ or
$V_{B}$. Then, after a change of the integration variable from $s$
to $a$ according to $ds=\left(\frac{da}{ds}\right)^{-1}da$, we
obtain
    \begin{equation}\label{FLRWIntegralsF0}
    \begin{split}
        -4\pi^2\int \frac{ds}{C(s)}\left[-3a+a^3V_{A,B}\right]=\pm12\pi^{2}i\int a\left(\sqrt{1-\frac{a^{2}}{3}V_{A,B}}\right) da \\ = \pm \frac{12\pi^2i}{V_{A,B}}\left[1-\frac{V_{A,B}}{3}a^2\right]^{3/2} .
    \end{split}
    \end{equation}
Putting $a(\bar{s}\pm\delta s)=\bar{a}\pm\delta a$ we obtain
    \begin{multline}\label{FLRWIntegralsF1E}
    -4\pi^2i\int_{0}^{\bar{s}-\delta s}\frac{ds}{C(s)}\left[-3a+a^3V_{A,B}\right]=\mp \frac{12\pi^2}{V_{A,B}}\left\{\left[1-\frac{V_{A,B}}{3}(\bar{a}-\delta a)^2\right]^{3/2}\right. \\ \left. -\left[1-\frac{V_{A,B}}{3}a_{0}^2\right]^{3/2}\right\} .
    \end{multline}
We note that this expression is well behaved in the limit
$a_{0}\to0$, therefore in the semiclassical approach we are working
on there is nothing that prevent us to safely choose the origin to
be at $a_{0}=a(s=0)=0$. Thus we finally obtain
    \begin{equation}\label{FLRWIntegralsF1}
        -4\pi^2i\int_{0}^{\bar{s}-\delta s}\frac{ds}{C(s)}\left[-3a+a^3V_{A,B}\right]=\mp \frac{12\pi^2}{V_{A,B}}\left\{\left[1-\frac{V_{A,B}}{3}(\bar{a}-\delta a)^2\right]^{3/2}-1\right\} .
    \end{equation}
In analogy with the computation of this type of transition
probabilities in the Euclidean regime, we can define a tension $T$
as the contribution to the action coming from the portion of the
path in which the scalar field is not a constant in the following
way
    \begin{equation}\label{FLRWTension}
        2\pi^2\bar{a}^3T=-4\pi^2i\int_{\bar{s}-\delta s}^{\bar{s}+\delta s}\frac{ds}{C(s)}a^3\left[V(\phi)-V_{A}\right]
        .
    \end{equation}
Then, substituting back the above equation (\ref{FLRWTension}) into
(\ref{FLRWGammaA}) we obtain the decay rate
$$
\pm\Gamma=\mp
\frac{12\pi^2}{\hbar}\left\{\frac{1}{V_{B}}\left[\left(1-\frac{V_{B}}{3}(\bar{a}-\delta
a)^2\right)^{3/2}-1\right]-\frac{1}{V_{A}}\left[\left(1-\frac{V_{A}}{3}(\bar{a}-\delta
a)^2\right)^{3/2}-1\right]\right\}
$$
\begin{equation}\label{FLRWGamma}
        +\frac{2\pi^2}{\hbar}\bar{a}^3T .
\end{equation}
Notice that the sign ambiguity $\pm$ in the left-hand side is not
related to the sign ambiguity $\mp$ on the right-hand side, we could
choose a sign in the right-hand and preserve the $\pm$ in the left
since they come from different arguments\footnote{This argument
holds for any equations involving two sign ambiguities in this
work.}. This result was derived in \cite{Cespedes:2020xpn} but by
proposing the value $C^2(s)=-1$ which we note that it is
inconsistent with the general system of equations (\ref{DefCs}) and
(\ref{RelDerC2}), but nevertheless gives the correct expression for
the classical action (\ref{FLRWIntegralsF0}).

We can recover the Euclidean result for the
tunneling probability when we apply the thin wall approximation,
that is when $\delta s\to0$, in this case we have
    \begin{equation}\label{FLRWGammaCDL}
    \pm\Gamma=\mp \frac{12\pi^2}{\hbar}\left\{\frac{1}{V_{B}}\left[\left(1-\frac{V_{B}}{3}\bar{a}^2\right)^{3/2}-1\right]-\frac{1}{V_{A}}\left[\left(1-\frac{V_{A}}{3}\bar{a}^2\right)^{3/2}-1\right]\right\}+\frac{2\pi^2}{\hbar}\bar{a}^3T .
    \end{equation}
Looking for the points that extremize the latter expression, that is
$\bar{a}$ such that $\frac{\partial\Gamma}{\partial\bar{a}}=0$, we
obtain
    \begin{equation}\label{FLRWExtrem}
        \frac{1}{\bar{a}^2}=\frac{V_{B}}{3}+\left(\frac{\Delta V}{3T}+\frac{T}{4}\right)^2=\frac{V_{A}}{3}+\left(\frac{\Delta V}{3T}-\frac{T}{4}\right)^2 ,
    \end{equation}
where  $\Delta V=V_{A}-V_{B}$. Substituting back the above value of
$\bar{a}$ into equation (\ref{FLRWGammaCDL}) and choosing the plus
sign in the right-hand side, we finally obtain
\begin{equation}\label{FLRWGammaF}
    \pm\Gamma=\frac{12\pi^2}{\hbar}\left\{\frac{\left[4(V_{A}-V_{B})^2+3T^2(V_{A}+V_{B})\right]\bar{a}}{12TV_{A}V_{B}}+\frac{1}{V_{A}}-\frac{1}{V_{B}}\right\} .
\end{equation}
These results coincide with the Euclidean computation for this case
presented in \cite{Parke:1982pm}. We note that this result is valid
for any nonzero values of $V_{A}$ and $V_{B}$, they could be
positive or negative. For the case in which one of the potential is
zero, the integral in (\ref{FLRWIntegralsF0}) would give as a result
a dependence of the form $\bar{a}^2$ and the following equations
must be changed accordingly.

We note that the expression for the logarithm of the
transition probability (\ref{FLRWGammaF}) is written in terms of two
factors, namely $\bar{a}$ and $T$. However, as we  have shown in
(\ref{FLRWExtrem}) they are not independent, therefore, effectively
the logarithm of the probability is described in this case in terms
of just one degree of freedom. Extremizing (\ref{FLRWGammaCDL}) we obtain
    \begin{equation}\label{FLRWTensionA}
        T=\pm 2\left(\sqrt{\frac{1}{\bar{a}^2}-\frac{V_{A}}{3}}-\sqrt{\frac{1}{\bar{a}^2}-\frac{V_{B}}{3}}\right) .
    \end{equation}
Then substituting it back in (\ref{FLRWGammaCDL})
and choosing the plus sign in the right-hand side we obtain
    \begin{multline}\label{FLRWGammaCDLF}
    \pm\Gamma= \frac{12\pi^2}{\hbar}\left\{\frac{1}{V_{B}}\left[\left(1-\frac{V_{B}}{3}\bar{a}^2\right)^{3/2}-1\right]-\frac{1}{V_{A}}\left[\left(1-\frac{V_{A}}{3}\bar{a}^2\right)^{3/2}-1\right]\right. \\ \left. +\frac{\bar{a}^2}{3}\left(\sqrt{1-\frac{V_{A}}{3}\bar{a}^2}-\sqrt{1-\frac{V_{B}}{3}\bar{a}^2}\right)\right\} .
    \end{multline}
Therefore, we note from (\ref{FLRWTensionA}) that in the limit
$\bar{a}\to0$ the tension vanishes. Consequently, the logarithm of
the transition probability  (\ref{FLRWGammaCDLF}) in this limit is
well behaved and it actually goes to zero. Therefore, the transition
probability goes to $1$ in the limit $\bar{a}\to0$. Thus, even if we
are working in a semiclassical approach in GR we can obtain a well
defined transition probability in the ultraviolet limit, so we think
our approach is worth studying despite its limited nature.

The transition probability just obtained depends on the possibility
to choose a path in field space such that the scalar field depends
on the parameter $s$ as in (\ref{ChooseS}). This choice was useful
to obtain the Euclidean results but we could also choose more
general paths. For instance, we could choose the parameter $s$ as
the distance in field space along the trajectories, that is we could
take
    \begin{equation}\label{ParameterSGeneral}
        ds^2=\int_{X}G_{MN}d\Phi^M\Phi^N .
    \end{equation}
By solving the resulting system we obtain that the general result
for the classical action (\ref{ClassActionGen}) takes the form
    \begin{equation}\label{FLRWClassicalActionGeneral}
        S_{0}=\pm 2\pi\int_{a_{0},\phi_{B}}^{a_{m},\phi_{A}}\frac{\left(a^3V(\phi)-3a\right)^{3/2}}{\sqrt{\frac{3}{2a}\left(a^2V(\phi)-1\right)^2-a^3(V'(\phi))^2}}\sqrt{a^3d\phi^2-6ada^2} ,
    \end{equation}
where prime denotes the derivative with respect to the scalar field.
We can see from the Hamiltonian constraint (\ref{FLREHam}) that the
last term, when taken as derivatives with respect to time, is
proportional to the kinetic part of the WDW equation as
    \begin{equation}\label{FLRWKinteicGeneral}
        a^3\dot{\phi}^2-6a\dot{a}^2=2N^2\left[\frac{\pi^2_{\phi}}{2a^3}-\frac{\pi^2_{a}}{12a}\right]=2N^2\left[3a-a^3V(\phi)\right] ,
    \end{equation}
and we note from (\ref{TransProb}) that the term contributing to the
transition probability is the imaginary part of the classical
action. Therefore, if the integrand of equation
(\ref{FLRWClassicalActionGeneral}) is real, then by using the last
expression we note that it is possible to obtain an imaginary
contribution to the action just by having the correct relations for
the momenta $\pi_{A}$ and $\pi_{\phi}$, that is, such that the last
expression is negative. This is possible in general because the
metric in superspace is not positive definite and therefore it is
possible to obtain a classical path.

\section{Transitions in a flat FLRW universe}
\label{S-FLRWP}

In this section we study the transitions in the case of a FLRW
metric with zero curvature. In this case we can write the metric in
cartesian coordinates as
    \begin{equation}\label{FLRWPMetric}
        ds^2=-N(t)dt^2+a^2(t)\left[dx^2+dy^2+dz^2\right] .
    \end{equation}
As usual in this case, the gravitational and scalar fields depend
only on the time variable $t$. Thus the corresponding Lagrangian is
given by
    \begin{equation}\label{FLRWPLagr}
    \mathcal{L}=-\frac{3a(t)\dot{a}(t)^2}{N(t)}+\frac{a^3(t)\dot{\phi}^2}{2N(t)}-N(t)a^3(t)V(\phi)
    .
    \end{equation}
The canonical momenta are the same as in Eq. (\ref{FLREMomenta}),
and the Hamiltonian constraint is
    \begin{equation}\label{FLRWPHam}
    H=N\left[\frac{\pi^2_{\phi}}{2a^3}-\frac{\pi^2_{a}}{12a}+a^3V\right]\approx 0
    .
    \end{equation}
As in the previous case we focus on the terms within brackets and
proceed to identify the metric in minisuperspace and the function
$f$. However, we can also factorize the Hamiltonian as follows
    \begin{equation}\label{FLRWPHamT}
    H=\frac{N}{a^3}\left[\frac{\pi^2_{\phi}}{2}-\frac{a^2\pi^2_{a}}{12}+a^6V\right]\approx 0 ,
    \end{equation}
and then focus on the terms within brackets. This choice is
analogous quantum mechanically to take a different ordering factor
to obtain the WDW equation. Comparing it with the general form
(\ref{HamConstraint}) we see that for this metric we have the
coordinates $\{\Phi^M\}=\{a,\phi\}$ and the nonzero elements of the
metric in minisuperspace are
    \begin{equation}\label{FLRWPMetricSuper}
    G^{aa}=-\frac{a^2}{6} , \hspace{1cm} G^{\phi\phi}=1 ,
    \end{equation}
and the function $f$ reads
    \begin{equation}\label{FLRWPf}
    f(a,\phi)=a^6V(\phi) .
    \end{equation}
We also note that in this case the spatial volume takes the form
    \begin{equation}\label{FLRWPVolume}
    {\rm Vol}(X)=\int\int\int dxdydz ,
    \end{equation}
which is finite only if we restrict ourselves to a finite interval
in the variables since in this case the spatial slice is not compact
and its volume might diverge. We note however, that since we
are working on minisuperspace, all fields depend only on time.
Therefore, we would consider a regularized version of the
probability transitions by taking an spatial section that is
appropriately compactified and which leads to a finite value for
this term. The corresponding probability transitions will be also
bounded from above and they will be finite and still might be useful
to describe these transitions for the non-compact cases. In any
case, the following procedure is valid without
changes\footnote{This argument holds for any volume term that does
not come from a compact spatial slice for the following sections of
this article.}

We consider the parameter $s$ defined as before in (\ref{ChooseS}),
then we have using Eq. (\ref{ClassActionGen}) for the classical
actions
\begin{multline}\label{FLRWPAccionComplA}
S_{0}(a_{0},\phi_{B};a_{m},\phi_{A})=-2{\rm
Vol}(X)\int_{0}^{s_{M}}\frac{ds}{C(s)}\left[a^6V(\phi)\right]=-2{\rm
Vol}(X)\left[\int_{0}^{\bar{s}-\delta
s}\frac{ds}{C(s)}\left[a^6V_{B}\right]\right. \\ \left. +
\int_{\bar{s}-\delta s}^{\bar{s}+\delta
s}\frac{ds}{C(s)}\left[a^6V(\phi)\right]+\int_{\bar{s}+\delta
s}^{s_{M}}\frac{ds}{C(s)}\left[a^6V_{A}\right]\right] ,
\end{multline}
\begin{equation}\label{FLRWPAccionComplB}
S_{0}(a_{0},\phi_{A};a_{m},\phi_{A})=-2{\rm
Vol}(X)\int_{0}^{s_{M}}\frac{ds}{C(s)}\left[a^6V_{A}\right] .
\end{equation}
Substituting these expressions into Eq. (\ref{DefGamma}) we obtain
the decay rate
\begin{multline}\label{FLRWPGammaA}
\pm\Gamma=-\frac{2{\rm Vol}(X)i}{\hbar}\int_{0}^{\bar{s}-\delta
s}\frac{ds}{C(s)}\left[a^6V_{B}\right]+\frac{2{\rm
Vol}(X)i}{\hbar}\int_{0}^{\bar{s}-\delta
s}\frac{ds}{C(s)}\left[a^6V_{A}\right]\\-\frac{2{\rm
Vol}(X)i}{\hbar}\int_{\bar{s}-\delta s}^{\bar{s}+\delta
s}\frac{ds}{C(s)}a^6\left[V(\phi)-V_{A}\right].
\end{multline}

For the case when the scalar field is constant, we use Eqs.
(\ref{DefCs}) and (\ref{RelDerC2}) to compute the first two
integrals of Eq. (\ref{FLRWPGammaA}) for the flat FLRW metric. The
general solutions (\ref{CsGeneral}) and (\ref{DerivSolGen}) take in
this case the form
\begin{equation}\label{FLRWPSolutions}
\begin{split}
C^{2}(s)&=12{\rm Vol}^2(X)a^6V_{A,B} ,
\\
\frac{da}{ds}&=\frac{1}{{\rm Vol}(X)}\frac{a}{6}.
\end{split}
\end{equation}
In terms of the variable $a(s)$ we obtain
\begin{equation}\label{FLRWPIntegralsF0}
\begin{split}
-2{\rm Vol}(X)\int \frac{ds}{C(s)}\left[a^6V_{A,B}\right]=\pm2{\rm
Vol}(X)\sqrt{\frac{V_{A,B}}{3}}a^3.
\end{split}
\end{equation}
We note that again we can safely choose the origin to be at $a_{0}=0$, then we get
    \begin{equation}\label{FLRWPIntegralsF0Aux}
    \begin{split}
    -2{\rm Vol}(X)\int_{0}^{\bar{s}-\delta s} \frac{ds}{C(s)}\left[a^6V_{A,B}\right]=\pm2{\rm Vol}(X)\sqrt{\frac{V_{A,B}}{3}}(\bar{a}-\delta a)^3.
    \end{split}
    \end{equation}
In this case we can also define a tension term $T$ as the
contribution coming from the portion of the path in which the scalar
field is not a constant as follows
    \begin{equation}\label{FLRWPTension}
    {\rm Vol}(X)\bar{a}^6T=-2{\rm Vol}(X)i\int_{\bar{s}-\delta s}^{\bar{s}+\delta s}\frac{ds}{C(s)}a^6\left[V(\phi)-V_{A}\right]
    .
    \end{equation}
Then, substituting the previous expression into (\ref{FLRWPGammaA})
and considering the thin wall limit of interest, that is $\delta
s\to0$, we obtain
    \begin{equation}\label{FLRWPGamma}
    \pm\Gamma=\pm\frac{2i{\rm Vol}(X)}{\sqrt{3}\hbar}\left(\sqrt{V_{B}}-\sqrt{V_{A}}\right)\bar{a}^3+\frac{{\rm Vol}(X)}{\hbar}\bar{a}^6T .
    \end{equation}
Looking for the points that extremize the latter expression, that is
$\frac{\partial\Gamma}{\partial\bar{a}}=0$, we obtain two possible
solutions
    \begin{equation}\label{FLRWPMinimumConditions}
    \bar{a}^3 =
    \begin{cases}
    0 ,\\
    \mp\frac{i}{\sqrt{3}T}\left(\sqrt{V_{B}}-\sqrt{{V_{A}}}\right) .
    \end{cases}
    \end{equation}
The first option of (\ref{FLRWPMinimumConditions}) implies the
vanishing of expression (\ref{FLRWPGamma}), so we focus on the
second line. Since $\bar{a}$ should be real, we find that in order
to have an extremum for $\Gamma$ different from zero, we must choose
both $V_{A}$ and $V_{B}$ to be negative, in which case we obtain
    \begin{equation}\label{FLRWPMinimumValue}
        \bar{a}^3=\pm\frac{1}{\sqrt{3}T}\left(\sqrt{|V_{B}|}-\sqrt{|V_{A}|}\right) ,
    \end{equation}
and substituting back into (\ref{FLRWPGamma}) we finally obtain
    \begin{equation}\label{FLRWPGammaF}
        \pm2{\rm Re}[\Gamma]=-\frac{2{\rm Vol}(X)}{3\hbar T}\left(\sqrt{|V_{B}|}-\sqrt{|V_{A}|}\right)^2 .
    \end{equation}

For non-negative values of $V_{A,B}$ we obtain in general
    \begin{equation}\label{FLRWPGammaGen}
    \pm2{\rm Re}[\Gamma]=\mp\frac{4{\rm Vol}(X)}{\sqrt{3}\hbar}\Im\left[\left(\sqrt{V_{B}}-\sqrt{V_{A}}\right)\right]\bar{a}^3+\frac{2{\rm Vol}(X)}{\hbar}\bar{a}^6T ,
    \end{equation}
but without an extremum. In particular, notice that for a transition
between two dS spaces values, the only term contributing is the
tension term.

We note also that in any of the possible cases, that is, if
we regard $T$ as an independent term or related to $\bar{a}$ as in
(\ref{FLRWPMinimumValue}) whenever it is possible, the logarithm of
the transition probability (\ref{FLRWPGammaGen}) is well behaved in
the ultraviolet limit $\bar{a}\to0$ giving once again a vanishing
value for $\Gamma$.

\section{Transitions in Kantowski-Sachs}\label{S-KS}
Now that we have studied the simplest cases of isotropic universes,
let us move on to the study of anisotropic metrics. Let us start
with the Kantowski-Sachs metric. Using the parametrization of this
metric used by Misner in \cite{Misner} we have
    \begin{equation}\label{MetricKSO}
    ds^{2}=-N^{2}(t)dt^{2}+e^{2\sqrt{3}\beta(t)}dr^{2}+e^{-2\sqrt{3}(\beta(t)+\Omega(t))}\left[d\theta^{2}+\sin^{2}(\theta)d\psi^{2}\right] ,
    \end{equation}
with $0\leq\theta\leq\pi$ and $0\leq\psi\leq2\pi$. Let us define for
simplicity the functions $\gamma(t)=e^{\sqrt{3}\beta(t)}$ and
$\sigma(t)=e^{-\sqrt{3}\Omega(t)}$, then the metric is written as
    \begin{equation}\label{MetricKS}
    ds^2=-N^2(t)dt^2+\gamma^2(t)dr^2+\frac{\sigma^2(t)}{\gamma^{2}(t)}\left[d\theta^2+\sin^2\theta d\psi^2\right] .
    \end{equation}
Considering again gravity coupling to a scalar field depending only
on the time variable $\phi(t)$ with potential $V(\phi)$ and using
natural units, the Lagrangian is
    \begin{equation}\label{KSLagrangian}
        \mathcal{L}=\frac{\sigma^2(t)}{\gamma^3(t)N(t)}\dot{\gamma}^2-\frac{\dot{\sigma}^2}{\gamma(t)N(t)}+N(t)\gamma(t)+\left[\frac{\dot{\phi}^2}{2N(t)}-N(t)V(\phi)\right]\frac{\sigma^2(t)}{\gamma(t)} .
    \end{equation}
The canonical momenta are
    \begin{equation}\label{KSMomenta}
        \pi_{N}=0 , \hspace{1cm} \pi_{\gamma}=\frac{2\sigma^2}{N\gamma^3}\dot{\gamma} , \hspace{1cm} \pi_{\sigma}=-\frac{2}{N\gamma}\dot{\sigma} , \hspace{1cm} \pi_{\phi}=\frac{\sigma^2}{N\gamma}\dot{\phi}
    \end{equation}
and the Hamiltonian constraint turns out to be
    \begin{equation}\label{KSHamilt}
    H=N\left[\frac{\gamma^3}{4\sigma^2}\pi^2_{\gamma}-\frac{\gamma}{4}\pi^2_{\sigma}+\frac{\gamma}{2\sigma^2}\pi^2_{\phi}+\frac{\sigma^2}{\gamma}V(\phi)-\gamma\right]\approx 0 .
    \end{equation}
Again, we can focus only on the term inside brackets, and comparing
it with the general form (\ref{HamConstraint}) we see that the set
of coordinates is $\{\Phi^M\}=\{\gamma,\sigma,\phi\}$, the
nonvanishing element on the metric in minisuperspace are
    \begin{equation}\label{KSMetricSup}
    G^{\gamma\gamma}=\frac{\gamma^{3}}{2\sigma^2} , \hspace{0.5cm} G^{\sigma\sigma}=-\frac{\gamma}{2} , \hspace{0.5cm} G^{\phi\phi}=\frac{\gamma}{\sigma^2} ,
    \end{equation}
and the function $f$ is given by
    \begin{equation}\label{KSf}
         f(\gamma,\sigma,\phi)=\frac{\sigma^2}{\gamma}V(\phi)-\gamma .
    \end{equation}
We also have in this case that the volume of $X$ reads
    \begin{equation}\label{KSVOlume}
        {\rm Vol}(X)=\int_{r}\int_{\theta=0}^{\pi}\int_{\psi=0}^{2\pi}\sin\theta drd\theta d\psi=4\pi\int dr ,
    \end{equation}
which is finite only if we restrict ourselves to a finite interval
on $r$, since in this case again the spatial slice is not compact.

We choose the parameter $s$ as before in Eq. (\ref{ChooseS}),
therefore by using (\ref{ClassActionGen}) we obtain in this case that the
classical actions are given by
    \begin{multline}\label{KSClassicalAction01}
    S_{0}(\gamma_{0},\sigma_{0},\phi_{B};\gamma_{m},\sigma_{m},\phi_{A})=-2{\rm Vol}(X)\bigg\{\int_{0}^{\bar{s}-\delta
    s}\frac{ds}{C(s)}\left[\frac{\sigma^2}{\gamma}V_{B}-\gamma\right]
    \\
    +\int_{\bar{s}-\delta s}^{\bar{s}+\delta s}\frac{ds}{C(s)}\left[\frac{\sigma^2}{\gamma}V(\phi)-\gamma\right] + \int_{\bar{s}+\delta s}^{s_{M}}\frac{ds}{C(s)}\left[\frac{\sigma^2}{\gamma}V_{A}-\gamma\right]\bigg\} ,
    \end{multline}
    \begin{equation}\label{KSCLassicalAction02}
    S_{0}(\gamma_{0},\sigma_{0},\phi_{A};\gamma_{m},\sigma_{m},\phi_{A})=-2{\rm Vol}(X)\int_{0}^{s_{M}}\frac{ds}{C(s)}\left[\frac{\sigma^2}{\gamma}V_{A}-\gamma\right] ,
    \end{equation}
and then substituting into (\ref{DefGamma}) we obtain the
corresponding decay rate
    \begin{multline}\label{KSGammaDef}
    \pm\Gamma=-\frac{2{\rm Vol}(X)i}{\hbar}\int_{0}^{\bar{s}-\delta s}\frac{ds}{C(s)}\left[\frac{\sigma^2}{\gamma}V_{B}-\gamma\right]+\frac{2{\rm Vol}(X)i}{\hbar}\int_{0}^{\bar{s}-\delta s}\frac{ds}{C(s)}\left[\frac{\sigma^2}{\gamma}V_{A}-\gamma\right] \\ -\frac{2{\rm Vol}(X)i}{\hbar}\int_{\bar{s}-\delta s}^{\bar{s}+\delta s}\frac{ds}{C(s)}\frac{\sigma^2}{\gamma}\left[V(\phi)-V_{A}\right] .
    \end{multline}
To compute the first two integrals we again study the system of
equations corresponding to this metric in the case in which the
scalar field is constant. The general solutions (\ref{CsGeneral})
and (\ref{DerivSolGen}) in this case read
    \begin{equation}\label{KSSolutionSystem}
    \begin{split}
    C^2(s)&={\rm Vol}^2(X)\left(\frac{\gamma^2}{\sigma^2}+3V_{A,B}\right) ,
    \\ \frac{d\gamma}{ds}&=\frac{1}{{\rm Vol}(X)}\frac{\gamma\left(\frac{\gamma^2}{\sigma^2}+V_{A,B}\right)}{\frac{\gamma^2}{\sigma^2}+3V_{A,B}} ,
    \\ \frac{d\sigma}{ds}&=\frac{1}{{\rm Vol}(X)}\frac{2\sigma V_{A,B}}{\frac{\gamma^2}{\sigma^2}+3V_{A,B}} .
    \end{split}
    \end{equation}
Now let us assume that neither $V_{A}$ nor $V_{B}$ are zero, then
the relations (\ref{FieldsRelations}) in this case give us
    \begin{equation}\label{KSRelationFields}
    \frac{1}{\gamma\left(\frac{\gamma^2}{\sigma^2}+V_{A,B}\right)}d\gamma=\frac{1}{2\sigma V_{A,B}}d\sigma ,
    \end{equation}
which can be written as the differential equation
    \begin{equation}
        2\sigma V_{A,B}d\gamma-\gamma\left(\frac{\gamma^2}{\sigma^2}+V_{A,B}\right)d\sigma=0
        .
    \end{equation}
This last expression leads to the relation
    \begin{equation}\label{KSRelationiFieldsComplete}
        \gamma^2=\frac{\sigma^2V_{A,B}}{1-c\sigma} ,
    \end{equation}
where $c$ is an integration constant. We note that by definition
$\gamma$ and $\sigma$ are positive functions, therefore the constant
$c$ has to fulfill the condition
    \begin{equation}\label{KSConditConst}
        \frac{V_{A,B}}{1-c\sigma}>0 ,
    \end{equation}
which depends on the value of $V_{A,B}$ and the possible values that
$\sigma$ can have. Having these two fields related, we can express
the first two integrals in (\ref{KSGammaDef}) in terms of $\sigma$
only. Moreover, changing the integration variable as
$ds=\left(\frac{d\sigma}{ds}\right)^{-1}d\sigma$ we can perform the
integral to obtain
    \begin{equation}\label{KSIntegralAu0}
    -2{\rm Vol}(X)\int\frac{ds}{C(s)}\left[\frac{\sigma^2}{\gamma}V_{A,B}-\gamma\right]=\pm {\rm Vol}(X)c_{A,B}F_{KS}[c_{A,B},\sigma] ,
    \end{equation}
    where we have defined the function
    $$
    F_{KS}[c,x]=\int\frac{\sqrt{4-3cx}}{1-cx}xdx
    $$
    \begin{equation}\label{FKS}
    =-\frac{2}{9c^2}\left[\sqrt{4-3cx}(5+3cx)-9\arctanh\left(\sqrt{4-3cx}\right)\right] .
    \end{equation}
In analogy with the FLRW metric (\ref{FLRWTension}), we can also
define now a tension term as
    \begin{equation}\label{KSTensionDef}
    {\rm Vol}(X)\frac{\bar{\sigma}^2}{\bar{\gamma}}T=-2{\rm Vol}(X)i\int_{\bar{s}-\delta s}^{\bar{s}+\delta s}\frac{ds}{C(s)}\frac{\sigma^2}{\gamma}\left[V(\phi)-V_{A}\right] ,
    \end{equation}
but as we have shown the minisuperspace coordinates $\gamma$ and
$\sigma$ can be related. In fact, using the relation obtained in the
case in which the scalar field is a constant
(\ref{KSRelationiFieldsComplete}) and the way in which the parameter
$s$ was defined in (\ref{ChooseS}), we have
    \begin{equation}\label{RelationFieldsGeneral}
    \gamma=
    \begin{cases}
    \sigma\sqrt{\frac{V_{B}}{1-c_{B}\sigma}} , & 0<s<\bar{s}-\delta s,\\
    \gamma(\sigma) , & \bar{s}-\delta s<s<\bar{s}+\delta s,\\
    \sigma\sqrt{\frac{V_{A}}{1-c_{AA}\sigma}} , & \bar{s}+\delta s<s<s_{M} ,
    \end{cases}
    \end{equation}
where $c_{AA}$ is a constant that satisfies the condition
(\ref{KSConditConst}) but it is in general different from $c_{A}$ in
(\ref{KSIntegralAu0}) since that constant emerges on a different
path. In the last expression $\gamma(\sigma)$ in the middle interval
can be found in principle by solving the system of equations for the
case in which the scalar field can vary. However, we note that in
the thin wall limit in which we are interested to work with, $\delta
s\to0$, the latter simplifies to
    \begin{equation}\label{RelationFieldsGeneralTWL}
    \gamma=
    \begin{cases}
    \sigma\sqrt{\frac{V_{B}}{1-c_{B}\sigma}} , & 0<s<\bar{s}-\delta s,\\
    \sigma\sqrt{\frac{V_{A}}{1-c_{AA}\sigma}} , & \bar{s}+\delta s<s<s_{M} ,
    \end{cases}
    \end{equation}
and we can impose continuity on $\gamma$ to  obtain
    \begin{equation}\label{KSRelationsTWL0}
    \bar{\gamma}=\bar{\sigma}\sqrt{\frac{V_{B}}{1-c_{B}\bar{\sigma}}}=\bar{\sigma}\sqrt{\frac{V_{A}}{1-c_{AA}\bar{\sigma}}}=\frac{V_{B}-V_{A}}{\sqrt{(c_{B}-c_{AA})(c_{B}V_{A}-c_{AA}V_{B})}} ,
    \end{equation}
    and
    \begin{equation}\label{KSRelationTWL1}
    \bar{\sigma}=\frac{V_{A}-V_{B}}{c_{B}V_{A}-c_{AA}V_{B}} .
    \end{equation}
Substituting the above results in (\ref{KSTensionDef}) and gathering
it and the result (\ref{KSIntegralAu0}) in (\ref{KSGammaDef}) we
then obtain in the thin wall limit
    \begin{equation}\label{KSGammaOK}
    \pm\Gamma=\pm \frac{{\rm Vol}(X)i}{\hbar}\left[c_{B}F_{KS}[c_{B},\sigma]\bigg\rvert^{\bar{\sigma}}_{\sigma_{0}}-c_{A}F_{KS}[c_{A},\sigma]\bigg\rvert^{\bar{\sigma}}_{\sigma_{0}}\right]+\frac{{\rm Vol}(X)}{\hbar}\frac{\bar{\sigma}T}{\sqrt{V_{B}}}\sqrt{1-c_{B}\bar{\sigma}}
    .
    \end{equation}
Therefore, in this limit we obtain for $\Gamma$ an expression that
only depends on two variables, $\bar{\sigma}$ and $T$, in complete
analogy to the FLRW cases. Note that in this case we can also safely
choose $\sigma_{0}=0$ since the above expression is well behaved in that limit. If we want to find an extremum of this
expression, that is $\bar{\sigma}$ such that $\frac{\partial
\gamma}{\partial \bar{\sigma}}=0$, we obtain that the following
equation has to be hold
    \begin{equation}\label{KSCoditionExt}
        \frac{2-3c_{B}\bar{\sigma}}{2\sqrt{V_{B}(1-c_{B}\bar{\sigma})}}T=\mp i\bar{\sigma}\left[\frac{c_{B}\sqrt{4-3c_{B}\bar{\sigma}}}{1-c_{B}\bar{\sigma}}-\frac{c_{A}\sqrt{4-3c_{A}\bar{\sigma}}}{1-c_{A}\bar{\sigma}}\right]
        .
    \end{equation}
We note in particular that for two positive values of the
potentials, the last equation have as only solution $T=0$ which is
not of interest to us, and therefore, there is not a minimum with a
nonzero tension in that case.

Since we expect the constants to be real numbers, we finally obtain
    \begin{multline}\label{KSGammaOKF}
    \pm2{\rm Re}[\Gamma]=\frac{2{\rm Vol}(X)}{\hbar}\left\{\mp \left[c_{B}\Im[F_{KS}[c_{B},\sigma]]\bigg\rvert^{\bar{\sigma}}_{\sigma_{0}}-c_{A}\Im[F_{KS}[c_{A},\sigma]]\bigg\rvert^{\bar{\sigma}}_{\sigma_{0}}\right]\right. \\ \left.+\frac{\bar{\sigma}T}{\sqrt{V_{B}}}\sqrt{1-c_{B}\bar{\sigma}}\right\}
    .
    \end{multline}
Note that this result and the latter equations are valid for any
values of the potentials different from zero. The only difference
between positive or negative potentials are the choices of the
integration constants so the condition (\ref{KSConditConst}) holds
in each particular case. Once again we find that if we
regard $T$ as an independent parameter for positive values of the
potentials, or related to $\bar{\sigma}$ as in
(\ref{KSCoditionExt}), the limit $\bar{\sigma}\to\sigma_{0}=0$ is
well behaved and it is a regular function in this limit. Eq.
(\ref{KSCoditionExt}) actually gives $T\to0$ in this limit, and in
general it yields a vanishing value for $\Gamma$, allowing us to
study the ultraviolet limit.

For the case in which the potential is zero, the general solution
(\ref{CsGeneral}) and (\ref{DerivSolGen}) gives
    \begin{equation}\label{KSSolutionGenC0}
    \begin{split}
        C^2(s)&={\rm Vol}^2(X)\frac{\gamma^2}{\bar{\sigma}^2} ,
        \\ \frac{d \gamma}{ds}&=\frac{\gamma}{{\rm Vol}(X)} ,
        \\ \frac{d\sigma}{ds}&=0 , \hspace{0.3cm} \therefore \hspace{0.3cm} \sigma=\bar{\sigma},
    \end{split}
    \end{equation}
then we obtain for this case
    \begin{equation}\label{KSIntegralAu0C0}
    -2{\rm Vol}(X)\int\frac{ds}{C(s)}\left[\frac{\sigma^2}{\gamma}V-\gamma\right]\bigg\rvert_{V=0}=\pm 2{\rm Vol}(X)\bar{\sigma}\ln\gamma .
    \end{equation}
Let us choose in the first instance $V_{A}=0$, then $V_{B}<0$, in
this case the path in which the system tunnels from one minimum to
the other, $\sigma$ starts with one value and evolves until it
reaches a constant $\bar{\sigma}$ and then it remains at that value.
For this case we obtain
    \begin{equation}\label{KSGammaOKC01}
    \pm\Gamma=\pm \frac{{\rm Vol}(X)i}{\hbar}\left[c_{B}F_{KS}[c_{B},\sigma]\bigg\rvert^{\bar{\sigma}}_{\sigma_{0}}-2\bar{\sigma}\ln{\gamma}\bigg\rvert^{\bar{\sigma}\sqrt{\frac{V_{B}}{1-c_{B}\bar{\sigma}}}}_{\sigma_{0}\sqrt{\frac{V_{B}}{1-c_{B}\sigma_{0}}}}\right]+\frac{{\rm Vol}(X)}{\hbar}\frac{\bar{\sigma}T}{\sqrt{V_{B}}}\sqrt{1-c_{B}\bar{\sigma}} ,
    \end{equation}
from this we note that we cannot choose $\sigma_{0}=0$ since there
is a divergence in the logarithm. The condition to have an extremum
leads to the equation
         $$
        \frac{2-3c_{\sigma}\bar{\sigma}}{2\sqrt{V_{B}(1-c_{\sigma}\bar{\sigma})}}T=\mp i \bigg\{\frac{\sqrt{4-3c_{\sigma}\bar{\sigma}}}{1-c_{\sigma}\bar{\sigma}}c_{\sigma}\bar{\sigma}
         $$
        \begin{equation}\label{KsMinimumConditionC01}
        -2\bigg[\ln\left(\bar{\sigma}\sqrt{\frac{V_{B}}{1-c_{\sigma}\bar{\sigma}}}\right)-\ln\left(\sigma_{0}\sqrt{\frac{V_{B}}{1-c_{\sigma}\sigma_{0}}}\right)
        +1-\frac{c_{\sigma}\bar{\sigma}}{2(1-c_{\sigma}\bar{\sigma})}\bigg]\bigg\} .
    \end{equation}

In the other possible case we choose $V_{B}=0$, therefore $V_{A}>0$.
In this case everything is expressed in terms of $\gamma$ instead of
$\sigma$, then we have
    \begin{equation}\label{KSRelationC02}
        \bar{\sigma}=\frac{\bar{\gamma}}{2V_{A}}\left[\sqrt{c_{A}^2\bar{\gamma}^2+4V_{A}}-c_{A}\bar{\gamma}\right] ,
    \end{equation}
and consequently
\small
    \begin{multline}\label{KSGammaOKC02}
    \pm\Gamma=\pm \frac{{\rm Vol}(X)i}{\hbar}\bigg\{\frac{\bar{\gamma}}{V_{A}}\left[\sqrt{c_{A}^2\bar{\gamma}^2+4V_{A}}
    -c_{A}\bar{\gamma}\right]\ln\gamma\bigg\rvert^{\bar{\gamma}}_{\gamma_{0}}\\
    -c_{A}F_{KS}\left[c_{A},\frac{\gamma}{2V_{A}}\left(\sqrt{c_{A}^2\gamma^2+4V_{A}}-c_{A}\gamma\right)\right]\bigg\rvert^{\bar{\gamma}}_{\gamma_{0}}\bigg\}+\frac{{\rm Vol}(X)}{\hbar}\frac{\bar{\gamma} T}{4V_{A}^2}\left(\sqrt{c_{A}^2\bar{\gamma}^2+4V_{A}}-c_{A}\bar{\gamma}\right)^2 ,
    \end{multline}
\normalsize
\noindent in this case we also note that $\gamma_{0}=0$
is not possible because it causes a divergence. The condition for
this expression to have an extremum is fulfilled if the following
equation holds
\small
    \begin{multline}\label{KSConditionExtremumC02}
        \frac{T}{4V_{A}^2}\left[6c_{A}^2\bar{\gamma}^2+4V_{A}-\frac{2c_{A}\bar{\gamma}}{\bar{d}}\left(3c^2_{A}\bar{\gamma}^2+8V_{A}\right)\right] \\
        =\mp i\left[-\frac{c_{A}}{V_{A}}\left(\frac{c^2_{A}\bar{\gamma}^2+2V_{A}}{\bar{d}}-c_{A}\bar{\gamma}\right)\frac{\sqrt{4-\frac{3c_{A}}{2V_{A}}\bar{\gamma}\bar{b}}}{1-\frac{c_{A}}{2V_{A}}\bar{\gamma}\bar{b}} +\frac{\bar{b}}{V_{A}} \right. \\ \left. +\frac{2}{V_{A}}\left(\frac{c^2_{A}\bar{\gamma}^2+2V_{A}}{\bar{d}}-c_{A}\bar{\gamma}\right)\left(\ln\bar{\gamma}-\ln\gamma_{0}\right) \right] ,
    \end{multline}
\normalsize
\noindent
where
    \begin{equation}\label{KSConditionExtremumC02Aux}
        d=\sqrt{c^2_{A}\bar{\gamma}^2+4V_{A}} , \hspace{0.5cm} \bar{b}=\bar{d}-c_{A}\bar{\gamma} .
    \end{equation}

As we mentioned in the final part of section \ref{S-FLRW}, we could also used the distance along the trajectories in field space as the parameter $s$ defined in (\ref{ParameterSGeneral}). In this case, after solving the system of equations we obtain for the classical action the general expression
    \begin{equation}\label{KSClassicalActionGeneral}
        S_{0}=\pm2\sqrt{{\rm Vol}(X)}\int_{\gamma_{0},\sigma_{0},\phi_{B}}^{\gamma_{m},\sigma_{m},\phi_{A}}\frac{\sigma^2\left(V(\phi)-\frac{\gamma^2}{\sigma^2}\right)^{3/2}}{\gamma\sqrt{\left(3V(\phi)+\frac{\gamma^2}{\sigma^2}\right)\left(V(\phi)-\frac{\gamma^2}{\sigma^2}\right)-2(V'(\phi))^2}}ds ,
    \end{equation}
where
    \begin{equation}\label{KSdsGeneral}
        ds=\sqrt{\frac{2\sigma^2}{\gamma^3}d\gamma^2-\frac{2}{\gamma}d\sigma^2+\frac{\sigma^2}{\gamma}d\phi^2} ,
    \end{equation}
and using the Hamiltonian constraint (\ref{KSHamilt}) we can write  the $ds$ term as
    \begin{equation}\label{KSRelationDiffGeneral}
    \frac{2\sigma^2}{\gamma^3}\dot{\gamma}^2-\frac{2}{\gamma}\dot{\sigma}^2+\frac{\sigma^2}{\gamma}\dot{\phi}^2=2N^2\left[\frac{\gamma^3}{4\sigma^2}\pi^2_{\gamma}-\frac{\gamma}{4}\pi^2_{\sigma}+\frac{\gamma}{2\sigma^2}\pi^2_{\phi}\right]=2N^2\left[\gamma-\frac{\sigma^2}{\gamma}V(\phi)\right] .
    \end{equation}
Therefore, if the integrand in (\ref{KSClassicalActionGeneral}) is real, we can have contributions for the transition probability just by having a negative kinetic term in the WDW equation. Again, this is possible because the metric in superspace is not positive definite, allowing again for a classical path. We note that in this case, since we have three terms, we have more possibilities to realize these paths. Since we have only one negative term coming from the momenta $\pi_{\sigma}$, it may seem from the first equality that we could have transitions even if the scalar field is not considered. However, if we remove the scalar field from (\ref{KSClassicalActionGeneral}) we obtain
    \begin{equation}\label{KSClassicalActionGeneralNS}
        S_{0}\bigg\rvert_{\phi=0,V=0}=\mp2\sqrt{{\rm Vol}(X)}\int_{\gamma_{0},\sigma_{0}}^{\gamma_{m},\sigma_{m}}\sigma\sqrt{\frac{2\sigma^2}{\gamma^3}d\gamma^2-\frac{2}{\gamma}d\sigma^2} ,
    \end{equation}
and from the second equality in (\ref{KSRelationDiffGeneral}) we note that the last term cannot be negative, since it depends only on $\gamma$ in this case. Therefore, if we remove the scalar field, we have no contributions for the transition probability.


\section{Transitions in Bianchi III}
\label{S-B3} Another important anisotropic metric is the Bianchi III
metric. In this section we study probability transitions for this
anisotropic kind of models. The metric is of the form
\cite{HamiltonianCosmology}
    \begin{equation}\label{MetricB3}
    ds^{2}=-N^{2}(t)dt^{2}+A^{2}(t)dx^{2}+B^{2}(t)e^{-2\alpha x}dy^{2}+C^{2}(t)dz^{2} ,
    \end{equation}
where $\alpha\neq0$ is a constant. We note that if we take
$A(t)=B(t)=C(t)$ and $\alpha=0$ then we recover the FLRW flat metric
(\ref{FLRWPMetric}) . Considering a scalar field coupled to gravity
with the metric (\ref{MetricB3}) we obtain the Lagrangian
    \begin{multline}\label{B3Lagrangian}
    \mathcal{L}=-\frac{1}{N(t)}\left(\dot{A}(t)\dot{B}(t)C(t)+\dot{A}(t)\dot{C}(t)B(t)+\dot{B}(t)\dot{C}(t)A(t)\right)-\frac{\alpha^{2}}{A(t)}N(t)B(t)C(t)\\ +\frac{\dot{\phi}^{2}}{2N(t)}A(t)B(t)C(t)-V(\phi)N(t)A(t)B(t)C(t)
    .
    \end{multline}
The canonical momenta are
    \begin{equation}\label{B3CanoncialMomenta}
    \begin{split}
    \pi_{N}=0 , \hspace{0.5cm} \pi_{A}=-\frac{1}{N}(\dot{B}C+\dot{C}B) , \hspace{0.5cm} &\pi_{B}=-\frac{1}{N}(\dot{A}C+\dot{C}A) , \\
    \pi_{C}=-\frac{1}{N}(\dot{A}B+\dot{B}A) , \hspace{0.5cm} &\pi_{\phi}=\frac{\dot{\phi}}{N}ABC
    \end{split}
    \end{equation}
and the Hamiltonian takes the form
    \begin{multline}\label{B3Ham0}
    H=N\left[\frac{A}{4BC}\pi_{A}^{2}+\frac{B}{4AC}\pi_{B}^{2}+\frac{C}{4AB}\pi_{C}^{2}-\frac{1}{2C}\pi_{A}\pi_{B}-\frac{1}{2B}\pi_{A}\pi_{C}-\frac{1}{2A}\pi_{B}\pi_{C} \right. \\ \left. +\frac{\pi_{\phi}^{2}}{2ABC}+\frac{\alpha^{2}BC}{A}+V(\phi)ABC\right] .
    \end{multline}
However, as we did with the FLRW flat metric, we find it more useful
to factorize the Hamiltonian as follows
    \begin{multline}\label{B3Ham}
    H= \frac{N}{ABC}\bigg\{\frac{A^2}{4}\pi^2_{A}-\frac{AB}{2}\pi_{A}\pi_{B}-\frac{AC}{2}\pi_{A}\pi_{C}+\frac{B^2}{4}\pi^2_{B}-\frac{BC}{2}\pi_{B}\pi_{C}+\frac{C^2}{4}\pi^2_{C}+\frac{1}{2}\pi^2_{\phi}  \\  + B^2C^2\alpha^2+A^2B^2C^2V(\phi) \bigg\} ,
    \end{multline}
and then focus on the terms within the brackets. Comparing with the
general form (\ref{HamConstraint}) we see that the coordinates are
$\{\Phi^M\}=\{A,B,C,\phi\}$. Moreover the metric in the
minisuperspace is written as
    \begin{equation}\label{B3MetricSuper}
    (G^ {MN})= \frac{1}{2}\left( \begin{array}{cccc}
    A^2 & -AB & -AC & 0 \\
    -AB & B^2 & -BC & 0 \\
    -AC & -BC & C^2 & 0  \\
    0 & 0  & 0 & 2   \end{array} \right) ,
    \end{equation}
and the function $f$ reads
    \begin{equation}\label{B3f}
        f(A,B,C,\phi)=B^2C^2\left(\alpha^2+A^2V(\phi)\right) .
    \end{equation}
We also note that in this case the volume of $X$ is given by
    \begin{equation}\label{B3Volume}
        {\rm Vol}(X)=\int\int\int e^{-\alpha x}dxdyxdz ,
    \end{equation}
which again is finite only if we restrict to a finite interval in
$y$ and $z$. This happens because the spatial slice is not compact
as was the case for the flat FLRW and Kantowski-Sachs metrics.

Take the parameter $s$ as in (\ref{ChooseS}) we obtain the classical
actions
    \begin{multline}\label{B3ClassActi01}
    S_{0}(A_{0},B_{0},C_{0},\phi_{B};A_{m},B_{m},C_{m},\phi_{A})=-2{\rm Vol}(X)\left[\int_{0}^{\bar{s}-\delta s}\frac{ds}{C(s)}\left[B^2C^2\alpha^2+A^2B^2C^2V_{B}\right] \right. \\ \left. +\int_{\bar{s}-\delta s}^{\bar{s}+\delta s}\frac{ds}{C(s)}\left[B^2C^2\alpha^2+A^2B^2C^2V(\phi)\right]  \int_{\bar{s}+\delta s}^{s_{M}}\frac{ds}{C(s)}\left[B^2C^2\alpha^2+A^2B^2C^2V_{A}\right]\right] ,
    \end{multline}
    \begin{equation}\label{B3ClassActi02}
    S_{0}(A_{0},B_{0},C_{0},\phi_{A};A_{m},B_{m},C_{m},\phi_{A})=-2{\rm Vol}(X)\int_{0}^{s_{M}}\frac{ds}{C(s)}\left[B^2C^2\alpha^2 +A^2B^2C^2V_{A}\right]  ,
    \end{equation}
and therefore
    \begin{multline}\label{B3GammaDef}
    \pm\Gamma=-\frac{2{\rm Vol}(X)i}{\hbar}\int_{0}^{\bar{s}-\delta s}\frac{ds}{C(s)}B^2C^2\left[\alpha^2+A^2V_{B}\right] \\ +\frac{2{\rm Vol}(X)i}{\hbar}\int_{0}^{\bar{s}-\delta s}\frac{ds}{C(s)}B^2C^2\left[\alpha^2+A^2V_{A}\right] \\ -\frac{2{\rm Vol}(X)i}{\hbar}\int_{\bar{s}-\delta s}^{\bar{s}+\delta s}\frac{ds}{C(s)}A^2B^2C^2\left[V(\phi)-V_{A}\right] .
    \end{multline}

Using the same procedure as in the previous examples, we have for
the case in which the scalar field is constant and the potential is
nonzero, the solution
    \begin{equation}\label{B3SolutionGen}
    \begin{split}
    C^2(s)&=4{\rm Vol}^2(X)A^2B^2C^2V_{A,B}\frac{4\alpha^2+3A^2V_{A,B}}{\alpha^2+A^2V_{A,B}} ,
    \\ \frac{dA}{ds}&=\frac{1}{{\rm Vol}(X)}\frac{1}{2AV_{A,B}}\frac{(2\alpha^2+A^2V_{A,B})(\alpha^2+A^2V_{A,B})}{4\alpha^2+3A^2V_{A,B}} ,
    \\ \frac{dB}{ds}&=\frac{1}{{\rm Vol}(X)}\frac{B}{2}\frac{\alpha^2+A^2V_{A,B}}{4\alpha^2+3A^2V_{A,B}} , \\
    \frac{dC}{ds}&=\frac{1}{{\rm Vol}(X)}\frac{C}{2}\frac{\alpha^2+A^2V_{A,B}}{4\alpha^2+3A^2V_{A,B}} .
    \end{split}
    \end{equation}
Moreover the coordinates are related as
    \begin{equation}\label{B3RelField}
    \frac{1}{C}\frac{dC}{ds}=\frac{1}{B}\frac{dB}{ds}=\frac{AV_{A,B}}{2\alpha^2+A^2V_{A,B}}\frac{dA}{ds} ,
    \end{equation}
from where we obtain
    \begin{equation}\label{B3FieldRelAux}
    B=b_{0}C=c_{0}\sqrt{\frac{2\alpha^2}{V_{A,B}}+A^2} ,
    \end{equation}
where $b_{0}$ and $c_{0}$ are integration constants. Since they are
only multiplicative constants, they can be absorbed by redefinitions
on the spatial variables, therefore, for the two first integrals in
(\ref{B3GammaDef}) we have respectively
    \begin{equation}\label{B3FieldRel}
    B=C=\sqrt{\frac{2\alpha^2}{V_{A,B}}+A^2} .
    \end{equation}
Notice that if we take $\alpha=0$, the latter equation tells us that
$A=B=C$, that is, in the semiclassical approach, the only condition
to recover the flat FLRW metric is $\alpha=0$. With these relations
we can express the two first integrals in (\ref{B3GammaDef}) in
terms of $A$ only and then by changing the integration variable as
$ds=\left(\frac{dA}{ds}\right)^{-1}dA$ we can perform the remaining
integral to obtain
    \begin{equation}\label{B3ClassiActionResult}
    -2{\rm Vol}(X)\int\frac{ds}{C(s)}B^2C^2\left[\alpha^2+A^2V_{A,B}\right]=\pm2{\rm Vol}(X)\sqrt{V_{A,B}}F_{III}[V_{A,B},A] ,
    \end{equation}
where we have defined the function
\small
    \begin{multline}\label{F3}
    F_{III}[V,x]=\int\sqrt{\left(3x^2+\frac{4\alpha^2}{V}\right)\left(x^2+\frac{\alpha^2}{V}\right)}dx\\ =\frac{1}{9\sqrt{\frac{4\alpha^4}{V^2}+\frac{7\alpha^2}{V}x^2+3x^4}}\left\{\frac{12\alpha^4}{V^2}x+\frac{21\alpha^2}{V}x^3+9x^5 \right. \\ \left. + \frac{2\alpha^3i}{V^{5/2}}\sqrt{x^2V+\alpha^2}\sqrt{3Vx^2+4\alpha^2}\left[F\left(i\arcsin h\left(\sqrt{\frac{Vx}{\alpha^2}}\right),\frac{3}{4}\right)\right.\right. \\ \left.\left. -7E\left(i\arcsin
    h\left(\sqrt{\frac{Vx}{\alpha^2}}\right),\frac{3}{4}\right)\right]\right\}.
    \end{multline}
\normalsize \noindent Here $F(x,m)$ and $E(x,m)$ are the elliptic
integrals of first and second kind respectively. Note that if we
choose $\alpha=0$ the integral in the last expression simplifies and
(\ref{B3ClassiActionResult}) takes the same form as in the flat FLRW
case (\ref{FLRWPIntegralsF0}). We can also define a tension term as
    \begin{equation}\label{B3TensionDef}
    {\rm Vol}(X)\bar{A}^2\bar{B}^2\bar{C}^2T=-2{\rm Vol}(X)i\int_{\bar{s}-\delta s}^{\bar{s}+\delta s}\frac{ds}{C(s)}A^2B^2C^2\left[V(\phi)-V_{A}\right] .
    \end{equation}
With the solutions found above for the intervals in which the scalar
field is constant, we obtain that in the thin wall limit we have
    \begin{equation}\label{B3RelationsCases}
    B=C=
    \begin{cases}
    \sqrt{\frac{2\alpha^2}{V_{B}}+A^2} , & 0<s<\bar{s} \\
    b\sqrt{\frac{2\alpha^2}{V_{A}}+A^2} , & \bar{s}<s<s_{M} ,
    \end{cases}
    \end{equation}
where $b$ is a constant of integration that cannot be absorbed by
redefinition of the space variables. Therefore, if we require
continuity of the fields $B$ and $C$ we get
    \begin{equation}\label{B3Continuity1}
    \bar{B}=\bar{C}=\sqrt{\frac{2\alpha^2}{V_{B}}+\bar{A}^2}=b\sqrt{\frac{2\alpha^2}{V_{A}}+\bar{A}^2}=\sqrt{2}\alpha b\sqrt{\left(\frac{1}{b^2-1}\right)\left(\frac{1}{V_{B}}-\frac{1}{V_{A}}\right)} ,
    \end{equation}
and
    \begin{equation}\label{B3Continuity2}
    \bar{A}^2=\frac{2\alpha^2}{b^2-1}\left(\frac{1}{V_{B}}-\frac{b^2}{V_{A}}\right) .
    \end{equation}

Thus, we obtain in the thin wall limit
\small
    \begin{multline}\label{B3GammaResult}
    \pm\Gamma=\pm\frac{2{\rm Vol}(X)i}{\hbar}\left[\sqrt{V_{B}}F_{III}[V_{B},A]\bigg\rvert^{\bar{A}}_{A_{0}}-\sqrt{V_{A}}F_{III}[V_{A},A]\bigg\rvert^{\bar{A}}_{A_{0}}\right]\\ +\frac{{\rm Vol}(X)}{\hbar}\bar{A}^2\left(\frac{2\alpha^2}{V_{B}}+\bar{A}^2\right)^2T  .
    \end{multline}
\normalsize

\noindent Therefore for this metric we can also find an expression
for $\Gamma$ in terms of only two variables, in this case $\bar{A}$
and $T$. Note that in this case we have $\lim_{x\to0}F_{III}[V,x]=0$
for any value of $V$, therefore we can choose $A_{0}=0$ in the case
of having both $V_{A}$ and $V_{B}$ positive (since the relations
(\ref{B3FieldRel}) hold with $B$ and $C$ being real functions).

Since the latter expression is valid for any nonzero $V_{A,B}$ we
obtain finally
    \begin{multline}\label{B3GammaResultF}
    \pm2{\rm Re}[\Gamma]=\mp\frac{4{\rm Vol}(X)}{\hbar}\left\{\Im\left[\sqrt{V_{B}}F_{III}[A,V_{B}]\right]\bigg\rvert^{\bar{A}}_{A_{0}}-\Im\left[\sqrt{V_{A}}F_{III}[A,V_{A}]\bigg\rvert^{\bar{A}}_{A_{0}}\right]\right\} \\+\frac{2{\rm Vol}(X)}{\hbar}\bar{A}^2\left(\frac{2\alpha^2}{V_{B}}+\bar{A}^2\right)^2T  .
    \end{multline}
As we have pointed out before, taking the limit $\alpha\to0$, this
result simplifies to the corresponding result for the flat FLRW
metric (\ref{FLRWPGamma}), therefore showing consistency of this
method.

Looking for an extremum for this expression, that is looking for the
points that satisfy $\frac{\partial\Gamma}{\partial\bar{A}}=0$, we
obtain that the following equation fulfils
    $$
    T\bar{A}\left(3\bar{A}^4+\frac{8\alpha^2}{V_{B}}\bar{A}^2+\frac{4\alpha^4}{V^2_{B}}\right)=\mp i\bigg[\sqrt{\left(3\bar{A}^2V_{B}+4\alpha^2\right)\left(\bar{A}^2+\frac{\alpha^2}{V_{B}}\right)}
    $$
    \begin{equation}\label{B3MinCondit}
    -\sqrt{\left(3\bar{A}^2V_{A }+4\alpha^2\right)\left(\bar{A}^2+\frac{\alpha^2}{V_{A}}\right)}\bigg]
    .
    \end{equation}
We note that for the case in which both potentials are positive,
there is only the solution $T=0$. Therefore, in this case we again
does not have a minimum with a nonzero tension term for this case.

In this case for positive values for both minima of the
potential, since $T$ is an independent parameter, we can study the
limit $\bar{A}\to0$ in (\ref{B3GammaResultF}), giving a vanishing
result for $\Gamma$. We note that for this metric this is the only
situation in which we can access this limit.

\begin{figure}[h]
\centering
\includegraphics[width=0.9\textwidth]{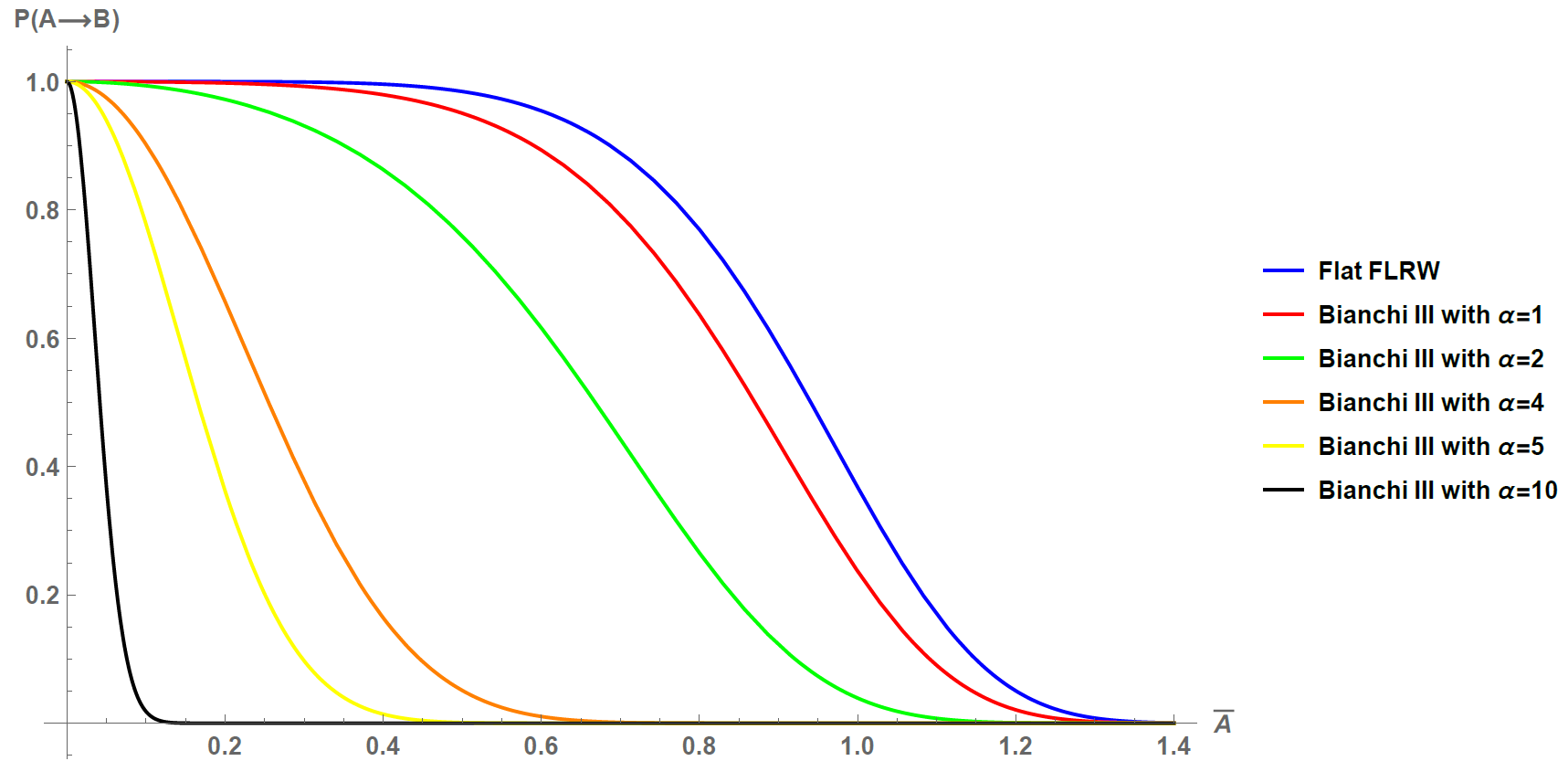}
\caption{Transition probability in units such that $\frac{2{\rm
Vol}(X)}{\hbar}=1$, with $V_{A}=100$, $V_{B}=10$ and $T=1$ for Bianchi
III with $\alpha=10$ (black line), $\alpha=5$ (yellow line), $\alpha=4$ (orange line),
$\alpha=2$ (green line), $\alpha=1$ (red line) and $\alpha=0$ or
flat FLRW (blue line). The contribution comes mostly from the tension term in (\ref{B3GammaResultF}) since the other term is about 16 order of magnitudes smaller in the whole range.} \label{FigTransProbN}
\end{figure}

For a value of the potential equal to zero, the system of equations
(\ref{DefCs}) and (\ref{RelDerC2}) for a constant scalar field is
inconsistent. Therefore, the results found for this metric are only
valid for $V_{A,B}\neq 0$. We could try another factorization of the
Hamiltonian or equivalently another ordering factor but we fail to
obtain a particular factorization that could lead us to both a
solvable system for a zero potential and an expression for the
integral in (\ref{B3ClassiActionResult}) that is computable.

\begin{figure}[h]
\centering
\includegraphics[width=0.9\textwidth]{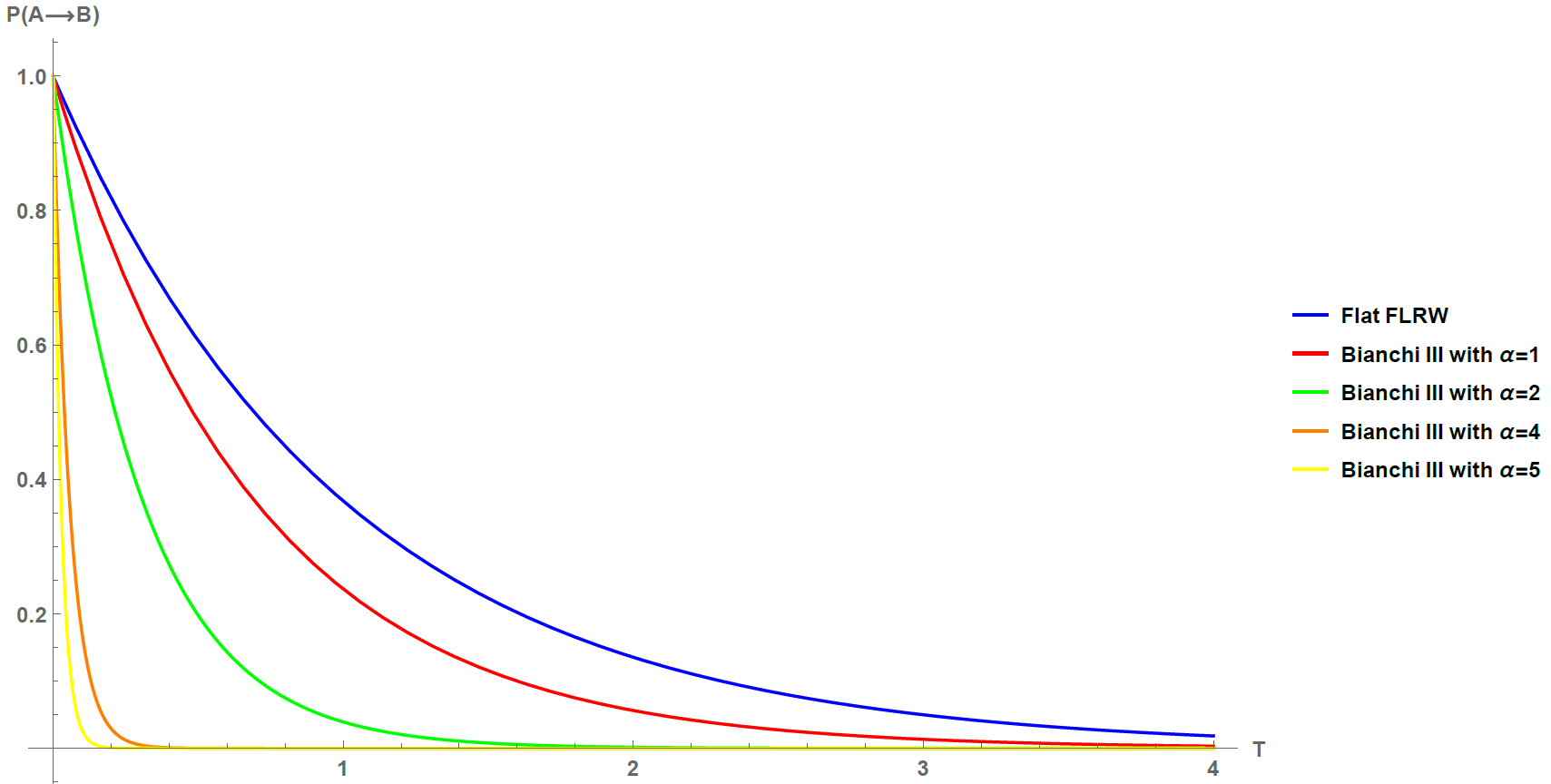}
\caption{Transition probability in units such that $\frac{2{\rm
Vol}(X)}{\hbar}=1$, with $V_{A}=100$, $V_{B}=10$ and $\bar{A}=1$ for
Bianchi III with $\alpha=5$ (yellow line), $\alpha=4$ (orange line),
$\alpha=2$ (green line), $\alpha=1$ (red line) and $\alpha=0$ or
flat FLRW (blue line).} \label{FigTransProbS}
\end{figure}

In order to visualize what is the consequence of having an
anisotropic universe for the transition probability, we plot the
probability (\ref{TransProb}) for this metric using
(\ref{B3GammaResultF}) with different values of $\alpha$ which
measures the degree of anisotropy of the system, having in the limit
$\alpha\to0$ the isotropy limit which is the flat FLRW case with
result (\ref{FLRWPGammaGen}). Let us choose the minus sign both in
the left and in the right of (\ref{B3GammaResultF}) so we have a
well behaved probability and the plot given in units such that
$\frac{2{\rm Vol}(X)}{\hbar}=1$ in both cases. We choose two
positive values for the potential minima, namely $V_{B}=10$,
$V_{A}=100$. Since both expressions don't have extrema, we firstly
set $T=1$ and plot the probability versus $\bar{A}$, or equivalently
$\bar{a}$. In Figure \ref{FigTransProbN} we present such plot, we
see that a generic feature of anisotropy is to reduce the transition
probability and make it fall faster as $\alpha^2$ increases. The
graph corresponding to isotropy, or the flat FLRW case, comes only
from the tension term in (\ref{FLRWPGammaGen}) as was explained
before, whereas the contribution coming from the Bianchi III cases
(\ref{B3GammaResultF}) come mostly from the tension term also since
for these values of the potential minima the contribution from the
other term is around 16 orders of magnitude smaller in the whole
range we are plotting. However, choosing different values of the
parameters does not change the behavior just described. Since we
have two independent variables in (\ref{B3GammaResultF}) we also
plot the probability versus the tension. Choosing the same
parameters as before and $\bar{A}=1$, we present this plot in Figure
\ref{FigTransProbS}. We see that the same behavior is encountered,
namely, the probability is decreased and falls faster as $\alpha^2$
increases.

In order to study transitions more generally, we can also used the
parameter $s$ defined in (\ref{ParameterSGeneral}). In this case we
obtain for the classical action
    \begin{equation}\label{B3ActionClassicalGeneral}
        S_{0}=\pm\sqrt{2{\rm Vol}(X)}\int_{A_{0},B_{0},C_{0},\phi_{B}}^{A_{m},B_{m},C_{m},\phi_{A}}\frac{BC\left(\alpha^2+A^2V(\phi)\right)^{3/2}}{A\sqrt{2V(\phi)\left(4\alpha^2+3A^2V(\phi)\right)-A^2(V'(\phi))^2}} ds ,
    \end{equation}
where
    \begin{equation}\label{B3dsGeneral}
        ds=\sqrt{-\frac{2}{AB}dAdB-\frac{2}{AC}dAdC-\frac{2}{BC}dBdC+d\phi^2} .
    \end{equation}
Using the Hamiltonian constraint with (\ref{B3Ham}) for this metric
we obtain that the $ds$ term can be related to the kinetic term in
the WDW equation as
    \begin{multline}\label{B3RelationsKinetic}
        -\frac{2}{AB}\dot{A}\dot{B}-\frac{2}{AC}\dot{A}\dot{C}-\frac{2}{BC}\dot{B}\dot{C}+\dot{\phi}^2\\=\frac{2N^2}{A^2B^2C^2}\left[\frac{A^2}{4}\pi^2_{A}-\frac{AB}{2}\pi_{A}\pi_{B}-\frac{AC}{2}\pi_{A}\pi_{C}+\frac{B^2}{4}\pi^2_{B}-\frac{BC}{2}\pi_{B}\pi_{C}+\frac{C^2}{4}\pi^2_{C}+\frac{1}{2}\pi^2_{\phi}\right] \\ =-\frac{2N^2}{A^2}\left(\alpha^2+A^2V(\phi)\right) .
    \end{multline}
Therefore, if the integrand in (\ref{B3ActionClassicalGeneral}) is
real, we can have contributions for the transition probability just
by having a negative kinetic term in the WDW equation, which can be
achieved by having correct values for the momenta, or as the second
equality of the last expression implies, just by having a positive
potential. If we want to explore a system without an scalar field,
the expression for the classical action does not have a well-defined
behavior, this happens because, as we have seen previously, the
system of differential equations in this case is inconsistent.

\section{Transitions in Bianchi IX}\label{S-B9}
In this subsection we finally study the anisotropic universe
described by the biaxial Bianchi IX metric
\cite{HamiltonianCosmology} given by
    \begin{equation}\label{B9MetricDefOr}
    ds^2=-\frac{N^2(t)}{q(t)}dr^2+\frac{p(t)}{4}\left(\sigma^2_{1}+\sigma^2_{2}\right)+\frac{q(t)}{4}\sigma^2_{3} ,
    \end{equation}
where
    \begin{equation}\label{B9MetricDefOrA}
    \begin{split}
    \sigma_{1}&=\sin\psi d\theta-\cos\psi\sin\theta d\omega ,
    \\ \sigma_{2}&=\cos\psi d\theta+\sin\psi\sin\theta d\omega , \\ \sigma_{3}&=-\left(d\psi+\cos\theta d\omega\right) ,
    \end{split}
    \end{equation}
and $0\leq\psi\leq4\pi$, $0\leq\theta\leq\pi$ and
$0\leq\omega\leq2\pi$.  For our purposes we write down the latter
expression for the metric (\ref{B9MetricDefOr}) in terms of the
trigonometric functions we have
    \begin{equation}\label{B9Metric}
    ds^2=-\frac{N^2(t)}{q}dt^2+\frac{q(t)}{4}d\psi^2+\frac{q(t)}{2}\cos\theta d\psi d\omega+\frac{p(t)}{4}d\theta^2+\frac{1}{4}\left(p(t)\sin^2\theta+q(t)\cos^2\theta\right)d\omega^2 .
    \end{equation}
Now we consider the presence of a scalar field as in the previous
cases, we have in this case the Lagrangian
    \begin{equation}\label{B9LagrangianDef}
        \mathcal{L}=\frac{1}{16}\left[-\frac{q}{2Np}\dot{p}^2-\frac{\dot{q}\dot{p}}{N}-\frac{2q}{p}N+8N\right]+\frac{p}{8}\left[\frac{q\dot{\phi}^2}{2N}-NV\right].
    \end{equation}
Moreover the canonical momenta are
    \begin{equation}
        \pi_{N}=0 , \hspace{0.5cm} \pi_{p}=-\frac{1}{16N}\left(\frac{q}{p}\dot{p}+\dot{q}\right) , \hspace{0.5cm} \pi_{q}=-\frac{\dot{p}}{16N} , \hspace{0.5cm}
        \pi_{\phi}=\frac{pq}{8N}\dot{\phi}.
    \end{equation}
Thus we obtain the Hamiltonian constraint
    \begin{equation}\label{B9HamiltConstra}
    H=N\left[\frac{8q}{p}\pi^2_{q}-16\pi_{p}\pi_{q}+\frac{4}{pq}\pi^2_{\phi}+\frac{1}{8}\left(\frac{q}{p}+pV(\phi)-4\right)\right] \approx 0 .
    \end{equation}
Focusing only in the term inside brackets in the previous equation
we have that in this case the coordinates in minisuperspace are
$\{\Phi^M\}=\{q,p,\phi\}$, the nonvanishing components of the
metric are
    \begin{equation}\label{B9MetricSuper}
    G^{qq}=\frac{16q}{p} , \hspace{0.5cm} G^{pq}=-16 , \hspace{0.5cm} G^{\phi\phi}=\frac{8}{pq} ,
    \end{equation}
and the function $f$ is given by
    \begin{equation}\label{B9fDef}
        f(p,q,\phi)=\frac{1}{8}\left(\frac{q}{p}+pV(\phi)-4\right) .
    \end{equation}
We also have in this case the volume of the three dimensional slice
is written as
    \begin{equation}\label{B9Volume}
        {\rm Vol}(X)=\int_{\omega=0}^{2\pi}\int_{\theta=0}^{\pi}\int_{\psi=0}^{4\pi}\sin\theta d\psi d\theta d\omega = 16\pi^2 ,
    \end{equation}
which is finite because the spatial slice considered is compact.

Moreover we choose the parameter $s$ as in the previous cases, then
we obtain for the classical actions
    \begin{multline}\label{B9ClassicalActionAu1}
    S_{0}(p_{0},q_{0},\phi_{B};p_{m},q_{m},\phi_{A})=-4\pi^2\bigg\{\int_{0}^{\bar{s}-\delta
    s}\frac{ds}{C(s)}\left[\frac{q}{p}+pV_{B}-4\right] \\
    +\int_{\bar{s}-\delta s}^{\bar{s}+\delta s}\frac{ds}{C(s)}\left[\frac{q}{p}+pV(\phi)-4\right] + \int_{\bar{s}+\delta s}^{s_{M}}\frac{ds}{C(s)}\left[\frac{q}{p}+pV_{A}-4\right]\bigg\} ,
    \end{multline}
    \begin{equation}\label{B9ClassicalActionAu2}
    S_{0}(p_{0},q_{0},\phi_{A};p_{m},q_{m},\phi_{A})=-4\pi^2\int_{0}^{s_{M}}\frac{ds}{C(s)}\left[\frac{q}{p}+pV_{A}-4\right].
    \end{equation}
Consequently then we have
    \begin{multline}\label{B9GammaDefO}
    \pm\Gamma=-\frac{4\pi^2i}{\hbar}\int_{0}^{\bar{s}-\delta s}\frac{ds}{C(s)}\left[\frac{q}{p}+pV_{B}-4\right]+\frac{4\pi^2i}{\hbar}\int_{0}^{\bar{s}-\delta s}\frac{ds}{C(s)}\left[\frac{q}{p}+pV_{A}-4\right] \\ -\frac{4\pi^2i}{\hbar}\int_{\bar{s}-\delta s}^{\bar{s}+\delta s}\frac{ds}{C(s)}p\left[V(\phi)-V_{A}\right] .
    \end{multline}

In the case of a constant scalar field we obtain the solution
    \begin{equation}\label{B9SolSyst}
    \begin{split}
    C^2(s)&=-1024\pi^4\frac{3q-2V_{A,B}p^2}{p^2\left(q+p^2V_{A,B}-4p\right)} ,\\ \frac{dq}{ds}&=\frac{1}{16\pi^2}\frac{(2q-V_{A,B}p^2)(q+p^2V_{A,B}-4p)}{3q-2V_{A,B}p^2} , \\ \frac{dp}{ds}&=-\frac{1}{16\pi^2}\frac{p(q+p^2V_{A,B}-4p)}{3q-2V_{A,B}p^2} ,
    \end{split}
    \end{equation}
whose fields are related as
    \begin{equation}\label{B9RelationFieldsOr}
    \frac{1}{2q-V_{A,B}p^2}\frac{dq}{ds}=-\frac{1}{p}\frac{dp}{ds} .
    \end{equation}
That expression can be rewritten as the differential equation
    \begin{equation}\label{B9RelationFieldsDE}
    pdq+(2q-V_{A,B}p^2)dp=0 ,
    \end{equation}
from where we obtain
    \begin{equation}\label{B0RelationFieldsT}
    q=\frac{c}{p^2}+\frac{V_{A,B}}{4}p^2 ,
    \end{equation}
where $c$ is an integration constant. Since by definition both $p$
and $q$ must be positive functions, we note that we can impose the
condition $\lim_{p\to0}q=0$ in the interval we are interested in for
positive values of the potential minima so we can have a spatial
singularity at the beginning of the path, then
    \begin{equation}\label{B9RelationFields}
    q=\frac{V_{A,B}}{4}p^2 .
    \end{equation}
With this result we can again express the two first integrals in
(\ref{B9GammaDefO}) in terms of an integral in $p$ as in the
previous cases obtaining
    \begin{equation}\label{B9ClassACtion}
    -4\pi^2\int\frac{ds}{C(s)}\left[\frac{q}{p}+pV_{A,B}-4\right]=\pm2\pi^{2}\sqrt{5V_{A,B}}F_{IX}[V_{A,B},p] ,
    \end{equation}
where we have defined

    \begin{multline}\label{F9}
    F_{IX}[V,x]= \int\sqrt{x\left(\frac{5V}{16}x-1\right)}dx \\ =\frac{1}{4\left(\frac{5V}{16}\right)^{3/2}\sqrt{x\left(\frac{5V}{16}x-1\right)}}
 \left[\sqrt{\frac{5V}{16}}x\left(1-\frac{15V}{16}x+\frac{25V^2}{128}x^2\right) \right. \\ \left.-\sqrt{x\left(\frac{5V}{16}x-1\right)}\ln\left[\frac{5V}{16}\sqrt{x}+\sqrt{\frac{5V}{16}}\sqrt{\frac{5V}{16}x-1}\right]\right] .
    \end{multline}

We can also define a tension term in this case as
    \begin{equation}\label{B9TensionDef}
    16\pi^2\bar{p}T=-4\pi^2i\int_{\bar{s}-\delta s}^{\bar{s}+\delta s}\frac{ds}{C(s)}p\left[V(\phi)-V_{A}\right] .
    \end{equation}
Therefore, we obtain in the thin wall limit an expression for the
decay rate
    \begin{equation}\label{B9GammaRes1}
    \pm\Gamma=\pm\frac{2\pi^2i\sqrt{5}}{\hbar}\bigg\{\sqrt{V_{B}}F_{IX}[V_{B},p]\bigg\rvert^{\bar{p}}_{p_{0}}-\sqrt{V_{A}}F_{IX}[V_{A},p]\bigg\rvert^{\bar{p}}_{p_{0}}\bigg\}+\frac{16\pi^2}{\hbar}\bar{p}T ,
    \end{equation}
and again it is expressed in terms of only two independent
variables, namely $\bar{p}$ and $T$. In this case we can safely
choose in any case $p_{0}=0$.  Therefore we obtain finally
    \begin{equation}\label{B9GammaRes}
    \pm2{\rm Re}[\Gamma]=\mp\frac{4\pi^2\sqrt{5}}{\hbar}\bigg\{\Im\left[\sqrt{V_{B}}F_{IX}[V_{B},p]\right]\bigg\rvert^{\bar{p}}_{0}-\Im\left[\sqrt{V_{A}}F_{IX}[V_{A},p]\right]\bigg\rvert^{\bar{p}}_{0}\bigg\}+\frac{32\pi^2}{\hbar}\bar{p}T ,
    \end{equation}
and the condition for an extremum leads to the equation
    \begin{equation}\label{B9CondtionMinimumGamma}
            8T=\mp \sqrt{5}i\left[\sqrt{V_{B}\bar{p}\left(\frac{5V_{B}}{16}\bar{p}-1\right)}-\sqrt{V_{A}\bar{p}\left(\frac{5V_{A}}{16}\bar{p}-1\right)}\right] .
    \end{equation}
Both results are valid for any values of the potential minima
$V_{A,B}$ greater than zero. For negative values the integration
constant cannot be set to zero and then the integral in the
expression (\ref{B9ClassACtion}) cannot be performed for an
arbitrary constant. We find that in this case we have in the
limit $\bar{p}\to0$ that $T\to0$ and  $\Gamma\to0$, so once again we
can observe that the transition probability near the primordial
singularity is regular.

In the case in which one of the potentials is zero, we have the
solution for a constant scalar field
    \begin{equation}\label{B9SolSystC0}
    \begin{split}
    C^2(s)&=-1024\pi^4\frac{3q}{p^2\left(q-4p\right)} ,\\ \frac{dq}{ds}&=\frac{1}{24\pi^2}(q-4p) , \\ \frac{dp}{ds}&=-\frac{1}{48\pi^2}\frac{p(q-4p)}{q} ,
    \end{split}
    \end{equation}
and the coordinates are related through the expression
    \begin{equation}\label{B9RelationFieldsC00}
        \frac{1}{q}\frac{dq}{ds}=-\frac{2}{p}\frac{dp}{ds} .
    \end{equation}
Then, absorbing a multiplicative constant by redefinition of the
spatial variables we get
    \begin{equation}\label{B9RelationFieldsC0}
        q=\frac{1}{p^2} .
    \end{equation}
With these relations we obtain
    \begin{equation}\label{B9ClaasicActionC0}
    -4\pi^2\int\frac{ds}{C(s)}\left[\frac{q}{p}+pV-4\right]\bigg\rvert_{V=0}=\pm2\sqrt{3}\pi^{2}F^{0}_{IX}(p) ,
    \end{equation}
    where we have defined
    \begin{multline}\label{F90}
    F^0_{IX}(x)= \int\frac{\sqrt{4x^3-1}}{x^3}dx \\
    =-\frac{\sqrt{4x^3-1}}{2x^2} +\frac{i2^{1/3}3^{3/4}}{\sqrt{4x^3-1}}\sqrt{(-1)^{5/6}(2^{2/3}x-1)}\sqrt{1+2^{2/3}x+2^{4/3}x^2}
    \\
    \times F\left(\arcsin\left(\frac{\sqrt{-(-1)^{5/6}-i2^{2/3}x}}{3^{1/4}}\right),(-1)^{1/3}\right) .
    \end{multline}
The only possible case now is $V_{B}=0$ and $V_{A}>0$. Then, we
obtain in the limit considered
    \begin{equation}\label{B9GammaRes1C01}
    \pm\Gamma=\pm\frac{2\pi^2i}{\hbar}\bigg\{\sqrt{3}F^{0}_{IX}(p)\bigg\rvert^{\bar{p}}_{p_{0}}-\sqrt{5V_{A}}F_{IX}[V_{A},p]\bigg\rvert^{\bar{p}}_{p_{0}}\bigg\}+\frac{16\pi^2}{\hbar}\bar{p}T
    .
    \end{equation}
We note that $F^{0}_{IX}(p)$ is divergent in the limit $p\to0$,
therefore in this case $p_{0}\neq0$. Thus finally
    \begin{equation}\label{B9GammaRes1C01F}
    \pm2{\rm Re}[\Gamma]=\mp\frac{4\pi^2}{\hbar}\bigg\{\sqrt{3}\Im[F^{0}_{IX}(p)]\bigg\rvert^{\bar{p}}_{p_{0}}-\sqrt{5V_{A}}\Im[F_{IX}[V_{A},p]]\bigg\rvert^{\bar{p}}_{p_{0}}\bigg\}+\frac{32\pi^2}{\hbar}\bar{p}T .
    \end{equation}
If we look for an extremum of the above expression we obtain that
the following equation must hold
    \begin{equation}\label{B9ConditionExtremumC01}
        8T=\mp i \left[\frac{\sqrt{3(4\bar{p}^3-1)}}{\bar{p}^3}-\sqrt{5V_{A}\bar{p}\left(\frac{5V_{A}}{16}\bar{p}-1\right)}\right]
        .
    \end{equation}
Since $T$ is real, the above condition leads to the equation
    \begin{equation}\label{B9ConditionExtremumC01Exp}
        4\bar{p}^3-1=0 , \hspace{0.5cm} 8T=\pm i\sqrt{5V_{A}\bar{p}\left(\frac{5V_{A}}{16}\bar{p}-1\right)}
        .
    \end{equation}
Therefore we find
    \begin{equation}\label{B9ConditionExtremumC01ExpA}
        \bar{p}=4^{-1/3} , \hspace{0.5cm} T=\frac{1}{8}\sqrt{\frac{5V_{A}}{4^{1/3}}\left(1-\frac{5V_{A}}{16(4^{1/3})}\right)} ,
    \end{equation}
and consequently we get the condition
$V_{A}<\frac{16(4^{1/3})}{5}\approx5.08$. Then substituting Eq.
(\ref{B9ConditionExtremumC01ExpA}) back into (\ref{B9GammaRes1C01F})
we finally obtain
    \begin{multline}\label{B9GammaRes1C01FF}
    \pm2{\rm Re}[\Gamma]=2\left\{\frac{2\pi^2}{\hbar}\left[\sqrt{3}\Im[F^{0}_{IX}(p)]\bigg\rvert^{4^{-1/3}}_{p_{0}}-\sqrt{5V_{A}}\Im[F_{IX}[V_{A},p]]\bigg\rvert^{4^{-1/3}}_{p_{0}}\right]\right. \\ \left. +\frac{\sqrt{5V_{A}}\pi^2}{\hbar}\sqrt{1-\frac{5V_{A}}{16(4^{1/3})}} \right\} .
    \end{multline}
If the last condition does not hold, then the expression for
$\pm\Gamma$ does not have a minimum with a nonvanishing tension
term.

Finally, to study transitions more generally, we can  use the parameter $s$ defined in (\ref{ParameterSGeneral}). In this case we obtain for the classical action
\begin{equation}\label{B9ActionClassicalGeneral}
S_{0}=\pm\frac{\pi}{2}\int_{q_{0},p_{0},\phi_{B}}^{q_{m},p_{m},\phi_{A}} \frac{\left(q+p^2V(\phi)-4p\right)^{3/2}}{\sqrt{2p^2V(\phi)-3q-\frac{p^4}{2q}(V'(\phi))^2}} ds ,
\end{equation}
where
\begin{equation}\label{B9dsGeneral}
ds=\sqrt{-\frac{1}{8}dqdp-\frac{q}{16p}dp^2+\frac{pq}{8}d\phi^2} .
\end{equation}
Using the Hamiltonian constraint  (\ref{B9HamiltConstra}) for this metric we obtain
\begin{equation}\label{B9RelationsKinetic}
    -\frac{\dot{q}\dot{p}}{8}-\frac{q}{16p}\dot{p}^2+\frac{pq}{8}\dot{\phi}^2=2N^2\left[\frac{8q}{p}\pi^2_{q}-16\pi_{p}\pi_{q}+\frac{4}{qp}\pi^2_{\phi}\right]=-\frac{N^2}{4}\left(\frac{q}{p}+pV(\phi)-4\right)
\end{equation}
Therefore, if the integrand in (\ref{B9ActionClassicalGeneral}) is real, we can have contributions for the transition probability just by having a negative kinetic term in the WDW equation. It may seem from the first equality that if we remove the scalar field, we could have a negative term contributing to the probability. However, if we remove the scalar field, we obtain the action
    \begin{equation}\label{B9ActionClassicalGeneralNS}
    S_{0}\bigg\rvert_{\phi=0,V=0}=\pm\frac{\pi}{2}\int_{q_{0},p_{0}}^{q_{m},p_{m}}(q-4p)\sqrt{\frac{4p-q}{3q}}\sqrt{-\frac{1}{8}dqdp-\frac{q}{16p}dp^2} ,
    \end{equation}
but from (\ref{B9RelationsKinetic}) we obtain
    \begin{equation}\label{B9RelationsKineticNS}
    -\frac{\dot{q}\dot{p}}{8}-\frac{q}{16p}\dot{p}^2=\frac{N^2}{4}\left(\frac{4p-q}{p}\right) .
    \end{equation}
We note from these last expressions that it is not possible to have an imaginary term coming from the action.

\section{Final Remarks}\label{S-Dis}
In the present article we have described a general procedure that
applies the WKB approximation to quantum gravity encoded in the WDW
equation with a Lorentzian perspective. This procedure was proposed
originally in \cite{Cespedes:2020xpn} in the computation of
transition probabilities for vacuum decay in the presence of a
scalar potential.

Moreover we have presented explicit computations of the transition
probabilities in the thin wall limit for three homogeneous but
anisotropic metrics, as well as for the isotropic metric FLRW with
both positive and null curvature for comparative purposes. For the
FLRW metric with positive curvature, we reproduced the same result
obtained in Ref. \cite{Cespedes:2020xpn}. However in our proposal we
have a slight variation which allow us to obtain a system of
equations which is solvable and it does not require to propose a
particular path in field space, which has the advantage of obtaining
an unique possible result.

For both anisotropic and isotropic metrics we found analytic
expressions for the logarithm of the transition probabilities, which
depends on the potential minima and two variables, namely the
tension term and one coming from the metric. For the anisotropic
metrics we found that a consequence of the semiclassical
approximation is that in all cases the metric degrees of freedoms
are related, therefore reducing to one the degree of freedom needed
to compute the classical action, this is similar to the procedure
used in \cite{Folomeev} in which the degrees of freedom are also
reduced in order to compute the classical action. The last step is
to look for an extremum of the logarithm of the transition
probability.

In the case of FLRW with positive curvature it was possible to do it
obtaining at the end the transition probability as a function of
only the potential minima and the scale factor. However for all other
metrics this step was found to be troublesome. Since the existence
of such a minimum was found to be in general highly dependent on the
specific values of the potential minima. Thus they all have an
additional variable in its final result.

For example, for the flat FLRW, Kantowski-Sachs and Bianchi III
metrics it was found that no minimum exists with a nonzero tension
if both potential minima are positive, that is for transition from
dS to dS.  For Bianchi IX with positive potential minima the
existence of a minimum depends completely on the specific values of
the potential in such points. For Bianchi IX with the true vacuum at
zero potential, it was found that the existence of a minimum leads
to a condition for the false vacuum potential and concrete values
for the tension term and the one coming for the metric, therefore
having one less term that FLRW with positive curvature.

For the Bianchi III metric we plot the transition probability as a
function of the degree of freedom coming from the metric, which was
one of the scale factors, as well as a function of the tension. We
see that we recover the flat FLRW metric result taking the isotropic
limit and we show that the degree of anisotropy leads in general to
a smaller probability and a faster decay in its plot as anisotropy
is increased. This result coincides completely with the general
behavior described in \cite{Mansouri} for the Bianchi I model using
a different approach.

Although in this work we used the WKB approximation based on
GR, we obtain that many of the expressions found for the transition
probabilities are well behaved and represented by regular functions
in the ultraviolet limit. Thus we consider them reliable physical
results and it is worth studying these processes in an early
universe in despite that GR is non-renormalizable. In particular, we
found that the probabilities for the FLRW metrics and the
Kantowski-Sachs metric are well behaved for all minima of the
potential different from zero. For the Bianchi III and Bianchi IX
metrics this was possible only for positive values of the minima of
the potential. In any possible case, all probabilities goes to $1$
in this limit. We interpret this result as follows: the closer we
are from the initial singularity the more probable these transitions
are to occur. The probability will decrease as soon as we move away
from the singularity. However, certainly in order to study the
behavior near the big bang singularity it is necessary to use a
gravity theory proposal with a better behavior near the singularity
than GR. It would be interesting to study theories of this type as
string theory or Horava-Lifshitz theory in order to check if the WKB
approximation to quantum gravity, proposed
\cite{deAlwis:2019dkc,Cespedes:2020xpn} and used in the present
article, can be implemented in these theories. At least for
Horava-Lifshitz theory \cite{Horava:2009uw} which is very similar to
GR in the infrared and whose Einstein-Hilbert action is corrected
with higher-order terms in the spatial curvature in the ultraviolet,
it seems to be possible to apply the mentioned WKB formalism. Some
of these considerations will be reported in a forthcoming work.

Finally it is worth mentioning that in the examples of anisotropic
metrics we considered here, the only one with a finite volume is the
Bianchi IX type.  Thus, in this case it is also valid the result
obtained in Ref. \cite{Cespedes:2020xpn}, which asserts that the
universe will remain closed after a transition for anisotropic
universes. Therefore in this context the transition also respects
the closeness of the spatial universe. Thus this certainly extends
the debate landscape/swampland corresponding to open/closed
universes
\cite{Freivogel:2005vv,Kleban:2012ph,Hartle:2013oda,Hawking:2017wrd}
to the anisotropic context.

 \vspace{1cm}
\centerline{\bf Acknowledgments} \vspace{.5cm} It is a pleasure to
thank Professor F. Quevedo for comments on the manuscript. D.
Mata-Pacheco would also like to thank CONACyT for grant No. 784617.




\begin{thebibliography}{99}

\bibitem{Arnowitt:1962hi}
R.~L.~Arnowitt, S.~Deser and C.~W.~Misner, ``The Dynamics of general
relativity,'' Gen. Rel. Grav. \textbf{40}, 1997-2027 (2008)
doi:10.1007/s10714-008-0661-1 [arXiv:gr-qc/0405109 [gr-qc]].

\bibitem{Wheeler} J.A. Wheeler, ``Superspace and the nature of quantum geometrodynamics,''
pp 615-724 of {\it Topics in Nonlinear Physics}, (ed) N.J. Zabusky,
Springer-Verlag NY, Inc., (1968).

\bibitem{DeWitt} B.S. DeWitt,``Quantum theory of gravity I, The canonical
theory,'' Phys. Rev. {\bf 160} (5) (1967) 1113.

\bibitem{ColemanFT1}
S.~R.~Coleman, ``The Fate of the False Vacuum. 1. Semiclassical
Theory,'' Phys. Rev. D \textbf{15}, 2929-2936 (1977) [erratum: Phys.
Rev. D \textbf{16}, 1248 (1977)] doi:10.1103/PhysRevD.16.1248


\bibitem{ColemanFT2}
C.~G.~Callan, Jr. and S.~R.~Coleman,
``The Fate of the False Vacuum.
2. First Quantum Corrections,'' Phys. Rev. D \textbf{16}, 1762-1768
(1977) doi:10.1103/PhysRevD.16.1762

\bibitem{ColemanDeLuccia}
S.~R.~Coleman and F.~De Luccia,
``Gravitational Effects on and of
Vacuum Decay,'' Phys. Rev. D \textbf{21}, 3305 (1980)
doi:10.1103/PhysRevD.21.3305

\bibitem{Parke:1982pm}
S.~J.~Parke,
``Gravity, the Decay of the False Vacuum and the New Inflationary Universe Scenario,''
Phys. Lett. B \textbf{121}, 313-315 (1983)
doi:10.1016/0370-2693(83)91376-X

\bibitem{FMP1}
W.~Fischler, D.~Morgan and J.~Polchinski,
``Quantum Nucleation of
False Vacuum Bubbles,'' Phys. Rev. D \textbf{41}, 2638 (1990)
doi:10.1103/PhysRevD.41.2638

\bibitem{FMP2}
W.~Fischler, D.~Morgan and J.~Polchinski,
``Quantization of False
Vacuum Bubbles: A Hamiltonian Treatment of Gravitational
Tunneling,'' Phys. Rev. D \textbf{42}, 4042-4055 (1990)
doi:10.1103/PhysRevD.42.4042

\bibitem{deAlwis:2019dkc}
S.~P.~De Alwis, F.~Muia, V.~Pasquarella and F.~Quevedo, ``Quantum
Transitions Between Minkowski and de Sitter Spacetimes,'' Fortsch.
Phys. \textbf{68}, no.9, 2000069 (2020) doi:10.1002/prop.202000069
[arXiv:1909.01975 [hep-th]].

\bibitem{Cespedes:2020xpn}
S.~Cespedes, S.~P.~de Alwis, F.~Muia and F.~Quevedo,
Phys. Rev. D \textbf{104}, no.2, 026013 (2021)
doi:10.1103/PhysRevD.104.026013 [arXiv:2011.13936 [hep-th]].

\bibitem{Saadeh:2016sak}
D.~Saadeh, S.~M.~Feeney, A.~Pontzen, H.~V.~Peiris and J.~D.~McEwen,
``How isotropic is the Universe?,'' Phys. Rev. Lett. \textbf{117},
no.13, 131302 (2016) doi:10.1103/PhysRevLett.117.131302
[arXiv:1605.07178 [astro-ph.CO]].

\bibitem{Colin:2018ghy}
J.~Colin, R.~Mohayaee, M.~Rameez and S.~Sarkar, ``Evidence for
anisotropy of cosmic acceleration,'' Astron. Astrophys.
\textbf{631}, L13 (2019) doi:10.1051/0004-6361/201936373
[arXiv:1808.04597 [astro-ph.CO]].

\bibitem{Soltis:2019ryf}
J.~Soltis, A.~Farahi, D.~Huterer and C.~M.~Liberato,
``Percent-Level Test of Isotropic Expansion Using Type Ia Supernovae,''
Phys. Rev. Lett. \textbf{122} (2019) no.9, 091301
doi:10.1103/PhysRevLett.122.091301
[arXiv:1902.07189 [astro-ph.CO]].

\bibitem{Migkas:2020fza}
K.~Migkas, G.~Schellenberger, T.~H.~Reiprich, F.~Pacaud,
M.~E.~Ramos-Ceja and L.~Lovisari, ``Probing cosmic isotropy with a
new x-ray galaxy cluster sample through the $L_{\text{X}}-T$ scaling
relation,'' Astron. Astrophys. \textbf{636}, A15 (2020)
doi:10.1051/0004-6361/201936602 [arXiv:2004.03305 [astro-ph.CO]].

\bibitem{Migkas:2021zdo}
K.~Migkas, F.~Pacaud, G.~Schellenberger, J.~Erler,
N.~T.~Nguyen-Dang, T.~H.~Reiprich, M.~E.~Ramos-Ceja and L.~Lovisari,
``Cosmological implications of the anisotropy of ten galaxy cluster
scaling relations,'' Astron. Astrophys. \textbf{649}, A151 (2021)
doi:10.1051/0004-6361/202140296 [arXiv:2103.13904 [astro-ph.CO]].

\bibitem{Krishnan:2021dyb}
C.~Krishnan, R.~Mohayaee, E.~\'O.~Colg\'ain, M.~M.~Sheikh-Jabbari
and L.~Yin, ``Does Hubble tension signal a breakdown in FLRW
cosmology?,'' Class. Quant. Grav. \textbf{38}, no.18, 184001 (2021)
doi:10.1088/1361-6382/ac1a81 [arXiv:2105.09790 [astro-ph.CO]].

\bibitem{Krishnan:2021jmh}
C.~Krishnan, R.~Mohayaee, E.~\'O Colg\'ain, M.~M.~Sheikh-Jabbari and
L.~Yin, ``Hints of FLRW Breakdown from Supernovae,''
[arXiv:2106.02532 [astro-ph.CO]].

\bibitem{DelCampo}
S.~del Campo and A.~Vilenkin,
``Tunneling Wave Function for
Anisotropic Universe,'' Phys. Lett. B \textbf{224}, 45-48 (1989)
doi:10.1016/0370-2693(89)91047-2

\bibitem{JensenRuback}
L.~G.~Jensen and P.~J.~Ruback,
``Bubble Formation in Anisotropic
Cosmologies,'' Nucl. Phys. B \textbf{325}, 660-686 (1989)
doi:10.1016/0550-3213(89)90502-6

\bibitem{Folomeev}
V.~N.~Folomeev, V.~T.~Gurovich and I.~V.~Tokareva,
``Generation of
the scalar field and anisotropy at quantum creation of the closed
universe,'' Nuovo Cim. B \textbf{115}, 1091-1100 (2000)
[arXiv:gr-qc/9912098 [gr-qc]].

\bibitem{Mansouri}
R.~Mansouri and M.~Mohazzab,
``Tunneling in anisotropic cosmological
models,'' Class. Quant. Grav. \textbf{10}, 1353-1359 (1993)
doi:10.1088/0264-9381/10/7/011 [arXiv:gr-qc/9211024 [gr-qc]].

\bibitem{Misner}
C.~W.~Misner, ``Minisuperspace,'' edited by J.R. Klauder, {\it Magic
Without Magic} John Archibald Wheeler, A collection in honor of his
60th birthday, San Francisco: W. H. Freeman, pages 441-473 (1972).

\bibitem{HamiltonianCosmology}
M. Ryan, {\it Hamiltonian cosmology}, Springer Verlag (1972).

\bibitem{Freivogel:2005vv}
B.~Freivogel, M.~Kleban, M.~Rodriguez Martinez and L.~Susskind,
``Observational consequences of a landscape,'' JHEP \textbf{03}, 039
(2006) doi:10.1088/1126-6708/2006/03/039 [arXiv:hep-th/0505232
[hep-th]].

\bibitem{Kleban:2012ph}
M.~Kleban and M.~Schillo, ``Spatial Curvature Falsifies Eternal
Inflation,'' JCAP \textbf{06}, 029 (2012)
doi:10.1088/1475-7516/2012/06/029 [arXiv:1202.5037 [astro-ph.CO]].

\bibitem{Hartle:2013oda}
J.~Hartle and T.~Hertog, ``Anthropic bounds on \ensuremath{\Lambda}
from the no-boundary quantum state,'' Phys. Rev. D \textbf{88},
no.12, 123516 (2013) doi:10.1103/PhysRevD.88.123516 [arXiv:1309.0493
[astro-ph.CO]].

\bibitem{Hawking:2017wrd}
S.~W.~Hawking and T.~Hertog, ``A Smooth Exit from Eternal
Inflation?,'' JHEP \textbf{04}, 147 (2018)
doi:10.1007/JHEP04(2018)147 [arXiv:1707.07702 [hep-th]].

\bibitem{Horava:2009uw}
  P.~Ho$\check{\rm r}$ava,
  ``Quantum Gravity at a Lifshitz Point,''
  Phys.\ Rev.\ D {\bf 79}, 084008 (2009)
  doi:10.1103/PhysRevD.79.084008
  [arXiv:0901.3775 [hep-th]].

\end{thebibliography}
\end{document}